\documentclass[review]{elsarticle} 
\usepackage{graphicx}
\usepackage{xcolor}
\usepackage{adjustbox}
\usepackage{chemformula}
\usepackage{multirow}
\usepackage{dcolumn}
\usepackage{bm}
\usepackage[mathlines]{lineno}
\usepackage[normalem]{ulem}
\usepackage{amsmath}
\usepackage{amssymb}
\usepackage{xcolor}
\usepackage[margin=1.0in]{geometry}

\journal{} 

\begin{document}
\begin{frontmatter}
	\title{Effects of bulk viscosity, heat capacity ratio and Prandtl number on the dispersion relationship of the compressible Navier-Stokes equation}
\author{Swagata Bhaumik$^{3}$\corref{cor1}}
\ead{swagatabhaumik@iitism.ac.in}
\author{Sawant Omkar Deepak$^{1}$}
\ead{omkars30.20dr0128@mech.iitism.ac.in}
\cortext[cor1]{Corresponding author \\\it{Email address}: swagatabhaumik@iitism.ac.in}
\address{$^{1,2,3}${Department Of Mechanical Engineering, IIT (ISM)
    Dhanbad, Dhanbad - 826004, India}}

\begin{abstract}
  Here, variation of the dispersion characteristics of 3D linearized compressible Navier-Stokes equation (NSE) with respect to bulk viscosity ratio ($\kappa/\mu$), specific heat ratio $\gamma$ and Prandtl number $Pr$ is presented. The 3D compressible NSE supports five type of waves, two vortical, one entropic and two acoustic modes. While the vortical and entropic modes are non-dispersive in nature, the acoustic modes are dispersive only up to a certain bifurcation wavenumber. The characteristics and variation of relative (with respect to the vortical mode) diffusion coefficient for entropic and acoustic modes and a specially designed dispersion function for acoustic modes with depressed wavenumber $\eta=KM/Re$ is presented. We have shown that these functions only depend on $\eta$, $\kappa/\mu$, $\gamma$ and Prandtl number $Pr$ of the flow. At lower wavenumber components, the deviation of the dispersion function from the inviscid and adiabatic case is proportional to $\eta^2$ at the leading order and the relative diffusion coefficients increase linearly with $\kappa/\mu$ and $\gamma$ while varying inversely with $Pr$. When the bulk viscosity ratio is increased, the shape and extent of the dispersion function is altered significantly and the change is more significant for higher wavenumber components. The relative diffusion coefficient for entropic and acoustic modes show contrasting variation with wavenumber depending upon $\kappa/\mu$, $\gamma$ and $Pr$. We show by solving linearized compressible NSE that relatively significant evolution and radiation of acoustic and/or entropic disturbances are noted when the bulk viscosity ratio is close to the corresponding critical value for which the bifurcation wavenumber is maximum. Based on this criterion, we have presented a empirical relation to obtain $\kappa/\mu$ depending upon $\gamma$ and $Pr$ which would indicate the range of bulk viscosity ratio for obtaining relatively significant disturbance evolution. 
\end{abstract}

    
\end{frontmatter}

\section{Introduction} \label{Intro}

Wave mechanics is a fundamental topic in various fields of applied and theoretical physics, like continuum mechanics, electro-magneto-dynamics, quantum mechanics, optics etc. or a combination of these \cite{landau2013,whitham}. As described in \cite{whitham}: ``wave is a recognizable signal that is transferred from one part of the medium to another with a recognizable velocity of propagation''. As it propagates, it may undergo distortion, amplitude variation and change in propagation velocity. A definitive class of waves may be classified as dispersive waves for which an analogous dispersion relation relating the wave's frequency and wavenumber can be derived by incorporating the properties of the medium in which it is propagating. This is a fundamental concept in wave-mechanics and is applicable to various wave phenomena \textit{i.e.}, electromagnetic waves \cite{landau2013}, acoustic waves \cite{whitham}, surface gravity and water waves \cite{lighthill2001} and waves in elastic media \cite{landau1986} and many more. A detailed description of dispersive system may be found in \cite{whitham,lighthill2001,landau2013}.

Here, we attempt to provide a theoretical description of the disturbance evolution characteristics in a compressible, isotropic Newtonian flow by illustrating the variation of its dispersion relation with respect to bulk viscosity ratio, specific heat ratio ($\gamma$) and Prandtl number ($Pr$) from low to high wavenumber range. These studies are important, particularly in the context of the determination of sound radiation and propagation in various gaseous environment. Most of the studies on aeroacoustics have been performed assuming zero-bulk viscosity (Stokes’ hypothesis) \cite{colonius1997,tam2006recent,moreau2022,lele2014,bailly2010} and also considering fixed values of $\gamma$ and $Pr$. These studies are important and have shed very interesting light on the sources and factors determining far-field acoustic radiation \cite{unnikrishnan2018,gonzalez2016}. However, there is a need to augment these by considering real-gas effects which generally has non-zero bulk viscosity\cite{graves1999} but also temperature or density varying $\gamma$ and Prandtl number $Pr$ (for example in the case of dense or strongly interacting gases). Similar studies by incorporating real-gas effects on radiation characteristics of acoustic/entropic disturbances may be important for high temperature gases (like that inside an internal-combustion engine or over- or under-expanded jets) or in case of different atmospheric conditions (like the planet Jupiter where the atmosphere is predominantly composed of $75\%$ hydrogen and $24\%$ helium or in Mars where the atmosphere is composed of $95\%$ Carbon dioxide). A good review of various recent numerical efforts (mostly 2015 onwards) of simulating high-speed flows involving shocks, turbulence and combustion incorporating bulk viscosity effects is presented in \cite{sharma2023} and is not repeated here.

The manuscript is organized as follows: In Sec.~\ref{Sec_i}, a theoretical description of the dispersion relation of the linearized compressible NSE is provided along with governing disturbance evolution equation for vortical, entropic and acoustic modes. The general nature of the dispersion equation and its effect on the dispersive and diffusive behavior of the modes are also illustrated. Section~\ref{Sec_ii} begins with the introduction of relative diffusion coefficients and the dispersion function and subsequently elaborates the variation of these variables with bulk viscosity ratio, specific heat ratio and Prandtl number. The section also presents results for linearized disturbance evolution while varying bulk viscosity ratio for a given initial condition. Additionally, it develops an empirical criterion on the bulk viscosity ratio for which relatively significant acoustic or entropic disturbance evolution is noted. Lastly, summary and conclusion is presented in Sec.~\ref{sec_iii}.

\section{The linearized 3D compressible Navier-Stokes equation} \label{Sec_i} 

The conservative form of the $3D$ compressible Navier-Stokes equations (NSE) in vectorial notation are given as

\begin{eqnarray}
\frac{\partial \tilde{\rho}}{\partial \tilde{t}}+ \tilde{\nabla} \cdot \left(\tilde{\rho} {\bf \tilde{v}}\right) = 0 \label{eq1} \\ 
\frac{\partial }{\partial t} \left(\tilde{\rho} {\bf \tilde{v}} \right) + \tilde{\nabla} \cdot \left( \tilde{\rho} {\bf \tilde{v}}{\bf \tilde{v}}\right) =
- \tilde{\nabla} \tilde{p} + \tilde{\nabla} \cdot \tilde{\bf \tau} \label{eq2} \\
\frac{\partial }{\partial t} \left(\tilde{\rho} \tilde{e}_t \right) + \tilde{\nabla} \cdot \left( \tilde{\rho} {\bf \tilde{v}} \tilde{h}_t \right) =
\tilde{\nabla} \cdot \left( \tilde{\bf \tau} \cdot {\bf \tilde{v}} \right) - \tilde{\nabla} \cdot {\bf \tilde{q}} \label{eq3} 
\end{eqnarray}

\noindent Equations~(\ref{eq1}-\ref{eq3}) are written in the dimensional form, where $\tilde{\rho}$ represents fluid density; $\tilde{T}$ represents fluid temperature; $p$ represents thermodynamic pressure;
${\bf \tilde{v}}$ represent fluid velocity $\tilde{e}_i$ and $\tilde{e}_t=\frac{1}{2} \left( {\bf \tilde{v}}\cdot {\bf \tilde{v}} \right) + \tilde{e}_i$ represent the specific internal and total energy, respectively; $h_i=\tilde{e}_i+\tilde{p}/\tilde{\rho}$ and
$\tilde{h}_t=\tilde{e}_t+\tilde{p}/\tilde{\rho}$ represent specific internal and total enthalpy, respectively; ${\bf \tilde{q}}$ represent the heat-flux and $\tilde{\bf \tau}$ represent the components of the viscous stress tensor. 
For an isotropic and Newtonian fluid obeying Fourier law of heat conduction, the symmetric stress tensor and the heat-flux are given as \cite{Kundu}  
\begin{eqnarray}
  \tilde{\bf \tau} = -\sigma D {\bf I} + 2\mu \tilde{\bf \epsilon}; \;\; \tilde{\bf \epsilon} = \frac{1}{2}\left( \tilde{\nabla} {\bf \tilde{v}}+\tilde{\nabla}{\bf \tilde{v}}^T\right) \;\; \text{and} \;\; 
  {\bf \tilde{q}} = - \hat{k} \tilde{\nabla} \tilde{T} \nonumber 
\end{eqnarray}        

\noindent Here, $\tilde{\bf \epsilon}$ and $D=\tilde{\nabla}\cdot {\bf \tilde{v}}$ is the symmetric strain-rate tensor and the volumetric dilatation rate, respectively. The variables, $\mu$, $\sigma$ and $\hat{k}$ represent the dynamic and the second coefficient of viscosity and the thermal
conductivity of the fluid, respectively. For a Newtonian and isotropic fluid the bulk coefficient of viscosity is obtained $\kappa=\left(-\sigma+2\mu/3\right)$ and according to the Stokes' hypothesis \cite{Kundu} $\sigma = 2\mu/3$ or $\kappa=0$. For a calorically perfect gas, which is assumed here, the the equation of state is given as $\tilde{p}=R\;\tilde{\rho} \; \tilde{T}$ while $\tilde{e}_i=c_v \tilde{T}$ and $h_i=c_p \tilde{T}$.
  Here, $R$, $c_v=R/(\gamma-1)$, $c_p=\gamma R/(\gamma-1)$ and $\gamma=c_p/c_v$ are the universal gas constant, specific heat at constant volume and constant pressure and ratio of specific heats, respectively. 

  Next, we consider the evolution of small perturbation by linearizing the Eqs.~(\ref{eq1}-\ref{eq3}) with respect to mean velocity
  ${\bf v_m}=(U_m,V_m,W_m)$, density $\rho_m$,
  pressure $p_m$ and temperature $T_m$. These mean variables are treated here to be truly constant and not functions of space and time. One can non-dimensionalize the resulting linearized
  equations by using $U=|{\bf v_m}|=\sqrt{U_m^2+V_m^2+W_m^2}$ as reference velocity scale; $\rho_m$ as reference density scale; $T_m$ as reference temperature scale;
  $p_m=R\rho_m T_m$ as the reference pressure scale; $L$ as reference length scale and $L/U$ as reference time scale. The resulting non-dimensional perturbation equations
  are given in the vector notation as

\begin{eqnarray}
\frac{\mathcal{D} \rho'}{\mathcal{D} t}+ \nabla \cdot {\bf v'} = 0 \label{eq7} \\
\frac{\mathcal{D} {\bf v'}}{\mathcal{D} t}+\frac{1}{\gamma M^2} \nabla p' = \frac{1}{Re} \nabla^2 {\bf v'} +
\frac{\left(1-\alpha\right)}{Re} \nabla \left(\nabla \cdot {\bf v'} \right) \label{eq7a}\\ 
\frac{\mathcal{D} T'}{\mathcal{D}t}+\left(\gamma -1 \right) \nabla \cdot {\bf v'} = \frac{\gamma}{Pr} \frac{1}{Re} \nabla^2 T' \label{eq8} 
\end{eqnarray} 

\noindent where $\alpha =\sigma/\mu=\left(2/3-\kappa/\mu\right)$ and $Re = {\mu}/{\rho_m U L}$, $Pr = {c_p \mu}/{\hat{k}}$ and $M = {U}/{\sqrt{\gamma R T_m}}$ is the Reynolds number, Prandtl number and the Mach number of the mean-flow, respectively. In Eqs.~(\ref{eq7}-\ref{eq8}), the operator $\mathcal{D}/\mathcal{D}t$ is defined as
${\mathcal{D}}/{\mathcal{D}t} = \left( {\partial }/{\partial t}+{\bf c} \cdot \nabla \right) $, where ${\bf c} = \left(c_x,c_y,c_z\right)$
such that $c_x=U_m/U$, $c_y=V_m/U$ and $c_z=W_m/U$.


One can also relate the disturbance pressure $p'$, density $\rho'$ and temperature $T'$ using the non-dimensional perturbed equation of state which for an ideal calorically perfect gas
is given as  
\begin{eqnarray}
p' = \left(\rho' + T'\right) \label{eq9}  
\end{eqnarray} 
\noindent  The non-dimensional perturbation entropy $s'$ and vorticity ${\bf \Omega'}$ are
given as $s'=T'-p'\left({\gamma-1}\right)/{\gamma}$ and ${\bf \Omega'}=\nabla\times {\bf v'}$, respectively.
Therefore, the evolution equation for $s'$ and ${\bf \Omega'}$ are given as 
\begin{eqnarray}
  \frac{\mathcal{D}s'}{\mathcal{D} t} & = & \frac{1}{Pr}\frac{1}{Re} \nabla^2 T' \label{eq9a} \\
\frac{\mathcal{D}{\bf \Omega'}}{\mathcal{D} t} & = & \frac{1}{Re} \nabla^2 {\bf \Omega'} \label{eq9b}    
\end{eqnarray} 
\noindent Hence, the linearized vorticity transport equation is essentially given by the 3D convection-diffusion equation. One notes that because of the linearization of the governing equations, nonlinear baroclinic source term are absent in
Eq.~(\ref{eq9b}).

\subsection{Physical dispersion relation of 3D-LNSE} \label{sec2p1}

We obtain the physical dispersion relation of 3D-LNSE given by Eqs.~(\ref{eq7}-\ref{eq8}) by denoting the physical variables
in terms of corresponding Fourier-Laplace amplitudes as 
\begin{eqnarray}
\left( \rho', {\bf v'}, T' \right)^T = \int \int \int \left(\hat{\xi}, \hat{\bf \Phi}, \hat{\theta} \right)^T 
e^{i\left(k_x x+k_y y+k_zz - \omega t\right)} dk_x dk_y dk_z \label{eq10} 
\end{eqnarray} 
\noindent where, $k_x$, $k_y$ and $k_z$ are wavenumbers along $x$-, $y$- and $z$-directions, respectively, while $\omega$ represents
the complex frequency and $i=\sqrt{-1}$. Note that $\omega=\omega_R+i\omega_I$ is a complex quantity such that $\omega_R$ denotes
the real frequency while $\omega_I$ denotes the temporal growth rate of the disturbances. On substituting Eq.~(\ref{eq10}) in
Eqs.~(\ref{eq7}-\ref{eq8}), the evolution eigenvalue equation is obtained as
$[{\bf D}-i\omega {\bf I}]\left( \hat{\xi}, \hat{\bf \Phi}, \hat{\theta} \right)^T=0$.
Hence, the physical dispersion relation is obtained by setting 
\begin{eqnarray}
|{\bf D}-i\omega {\bf I}|=\begin{vmatrix}
 \Lambda & ik_x & ik_y & ik_z & 0 \\
\frac{1}{\gamma M^2}ik_x & \Lambda+\Delta_1 & \Delta_5          & \Delta_6         & \frac{1}{\gamma M^2}ik_x \\
\frac{1}{\gamma M^2}ik_y & \Delta_5         & \Lambda+\Delta_2  & \Delta_7         & \frac{1}{\gamma M^2}ik_y \\
\frac{1}{\gamma M^2}ik_z & \Delta_6         & \Delta_7          & \Lambda+\Delta_3 & \frac{1}{\gamma M^2}ik_y \\
0 & \left(\gamma-1\right)ik_x & \left(\gamma-1\right)ik_y & \left(\gamma-1\right)ik_z & \Lambda+\Delta_4 \\
\end{vmatrix} = 0
\end{eqnarray}
\noindent where $\Lambda = i\left(\beta-\omega\right)$, $\beta=k_xc_x+k_yc_y+k_zc_z = {\bf K}\cdot {\bf c}$,
${\bf K}=\left(k_x,k_y,k_z\right)$, $\Delta_1=\left((2-\alpha)k_x^2+k_y^2+k_z^2\right)/Re$, 
$\Delta_2=\left(k_x^2+(2-\alpha)k_y^2+k_z^2\right)/Re$, $\Delta_2=\left(k_x^2+k_y^2+(2-\alpha)k_z^2\right)/Re$,
$\Delta_4=\gamma\left(k_x^2+k_y^2+k_z^2\right)/(Re Pr)$, $\Delta_5=(1-\alpha)k_x k_y/Re$, $\Delta_6=(1-\alpha)k_x k_z/Re$ and
$\Delta_7=(1-\alpha)k_y k_z/Re$. Here, ${\bf K}=\left(k_x,k_y,k_z\right)$ denotes the wavenumber vector whose modulus is given
as $K=\sqrt{k_x^2+k_y^2+k_z^2}$.

Simplifying the above relation, the characteristic equation is obtained as
\begin{eqnarray}
  & \Lambda^5 + C_1 z_1 \Lambda^4 + \left(C_2z_1^2+z_2 \right)\Lambda^3+ \left( C_3 z_1z_2+C_4z_1^3 \right) \Lambda^2 +
  \left(C_5 z_1^4+C_6z_1^2z_2\right)\Lambda+ \nonumber \\
  &  C_7z_1^3z_2 = 0 \label{eq12}  
\end{eqnarray}
\noindent where, $z_1=K^2/Re$, $z_2=\left(K/M\right)^2$, $K=\sqrt{k_x^2+k_y^2+k_z^2}$, $C_1=(4-\alpha)+\gamma/Pr$,
$C_2=(5-2\alpha)+(4-\alpha)\gamma/Pr$, $C_3=\left(2+1/Pr\right)$, $C_4=(2-\alpha)+(5-2\alpha)\gamma/Pr$, $C_5=(2-\alpha)\gamma/Pr$,
$C_6=(1+2/Pr)$, and $C_7=1/Pr$. It may noted that defining the depressed absolute wavenumber $\eta=\sqrt{z_1^2/z_2}=\left(KM/Re\right)$
and scaled dispersion variable $\lambda=\Lambda/z_1$, Eq.~(\ref{eq12}) may be expressed in terms of $\lambda$ as
\begin{eqnarray}
  & \lambda^5 + C_1 \lambda^4 + \left(C_2+\frac{1}{\eta^2} \right)\lambda^3+ \left( \frac{C_3}{\eta^2} +C_4 \right) \lambda^2 +
  \left(C_5 +\frac{C_6}{\eta^2} \right)\lambda+ \frac{C_7}{\eta^2} = 0 \label{eq12xy}  
\end{eqnarray}

\noindent Equation~(\ref{eq12}) is a quintic polynomial equation and generally solutions of any polynomial equation of degree five or higher can not be expressed in terms of radicals of the coefficients according to Abel-Ruffini theorem \cite{ramond2022}. However, this is not applicable here. One notes that the dispersion relationship can be expressed as $\omega_j = \beta + i\Lambda_j=\beta+iz_1\lambda_j$ where $j=1,2,3,4,5$ and segregating the real and imaginary parts, $\omega_{jR} = \beta - \Lambda_{jI}$ and $\omega_{jI} = \Lambda_{jR}$. It may be noted that the nondimensional number $M/Re$ appearing the expression for $\eta$, defines the ratio of the sound propagation time to the momentum diffusivity time \cite{ansumali2005}. This number may also be related to the Knudsen number $Kn=l/L$, where $l$ is the mean-free path of the fluid molecules. Following classical Maxwell-Boltzmann theory of a mono-atomic gas $M/Re=Kn\sqrt{2/(\pi \gamma)}$ \cite{sommerfeld1964}.

The $3D-LNSE$ is in general dispersive and diffusive in nature. For the inviscid and adiabatic
(and hence isentropic) case which corresponds to the linearized Euler equation (LEE), $C_1=C_2=C_3=C_4=C_5=C_6=C_7=0$ and therefore, 
$\omega_{1,2,3}=\beta$ while $\omega_{4,5} = \beta \mp K/M $. Therefore, the modes corresponding to $\omega_{1,2,3}$ are non-dispersive
and non-dissipative which only convects with the flow. As for the LEE, ${\mathcal{D}{\bf \Omega'}}/{\mathcal{D}t}={\mathcal{D}s'}/{\mathcal{D}t}=0$ and therefore, $\omega_{1,2,3}$ represent two vortical and one entropic modes, respectively. The modes denoted by $\omega_{4,5}$ are dispersive and non-diffusive. The propagation speed of these modes depends on the Mach number $M$ of the mean flow. These modes represent the two acoustic modes. This can be justified by considering the fact that following LEE, the acoustic equation corresponding to $p'$ is given as  
\begin{eqnarray}
\left(\frac{\mathcal{D}}{\mathcal{D}t}\right)^2p'-\frac{1}{M^2}\nabla^2p'=0 \label{eq12bc} 
\end{eqnarray} 
\noindent for which the dispersion relation is also given by $\omega = \beta\mp K/M$. For the present viscous case, $p'$ may be considered to be caused by acoustic and entropic disturbances simultaneously. Hence, using Eqs.~(\ref{eq7}) and (\ref{eq7a}) and the relationship $p'=\gamma \left( \rho'+ s' \right)$ that the modified acoustic relation may be obtained as   
\begin{eqnarray}
\left(\frac{\mathcal{D}}{\mathcal{D}t}\right)^2\rho'-\left(\frac{2-\alpha}{Re}\right) 
\left( \frac{\mathcal{D}}{\mathcal{D}t} \right) \nabla^2 \rho' -\frac{1}{M^2}\nabla^2 \rho'= \frac{1}{M^2}\nabla^2 s'
\label{eq15ab}
\end{eqnarray} 
\noindent Therefore, for the viscous case, the entropic perturbations act as a source term to the acoustic wave propagation
which are dissipative and dispersive in nature. In all subsequent discussions, modes-$1$ and $2$ represent two vortical modes;
mode-$3$ represents the entropic mode and modes-$4$ and $5$ represent the acoustic mode-$1$ and $2$, respectively.

It can be shown from the characteristic Eq.~(\ref{eq12}) that $\Lambda_{1,2}=-z_1=-K^2/Re$ are two roots of Eq.~(\ref{eq12}) and the
associated dispersion relation is given as $\omega_{1,2} = \beta-i{K^2}/{Re}$. These two modes correspond to the vortical mode as
can be readily noted from the perturbation linearized vorticity transport Eq.~(\ref{eq9b}). Factoring out these two roots from
Eq.~(\ref{eq12}), the dispersion relation for one entropic and the two acoustic modes are given by the following cubic equation 
\begin{eqnarray}
\Lambda^3 + C_8 z_1 \Lambda^2 + \left(C_5 z_1^2+z_2\right) \Lambda + C_7 z_1z_2 = 0 \label{eq12a}
\end{eqnarray} 
\noindent where $C_8=(2-\alpha)+\gamma/Pr$. In terms of $\lambda=\Lambda/z_1$, Eq.~(\ref{eq12a}) may be expressed as
\begin{eqnarray}
  \lambda^3 + C_8 \lambda^2 + \left(C_5 +{1}/{\eta^2} \right) \lambda + {C_7}/{\eta^2} = 0
  \label{eq12ab}
\end{eqnarray} 
\noindent It may be noted that for any general cubic equation, the roots may be obtained by applying the 
Cardano's method followed by Vieta's substitution \cite{cubic}. Similar dispersion analysis are also given in \cite{benjelloun2020} for $1D$-linearized compressible NSE. For $1D$ compressible NSE, considering heat-transfer one has only the entropic and the two acoustic modes and hence the dispersion relation is also given by a cubic equation \cite{benjelloun2020}.   

One notes that all the coefficients of Eq.~(\ref{eq12a}) are real, indicating that
it has either one real root ($\Lambda_3$) and two complex conjugate roots ($\Lambda_4$ and $\Lambda_5$) or three real roots. 
The real $\Lambda_3$ represents the entropic while the complex conjugate $\Lambda_4$ and $\Lambda_5$ denotes the acoustic modes. 
The analytical solution of a cubic polynomial equation is explained in Appendix-$I$ for the ease of the reader.


\subsection{Group velocity, phase velocity and diffusion coefficient}

For the above dispersive system, one can define the physical group velocity and the phase-speed of the $j^{th}$-mode
along $x$-direction as\cite{whitham}
\begin{eqnarray}
V_{gx}^j = \frac{\partial \omega_{jR}}{\partial k_x} = c_x - \frac{d \Lambda_{jI}}{dK} \frac{k_x}{K} \label{eq13a} \\ 
c_{px}^j = \frac{\omega_{jR}}{k_x} \label{eq13b}
  \label{group}
\end{eqnarray} 
\noindent The group velocity and the phase speed along $y$- and $z$-directions can also be similarly defined. For a system with
complex $\omega$, the physical wave solution grows in time as $e^{\omega_I t}$. This can be used to define the diffusion coefficient
of the $j^{th}$-mode as
\begin{equation}
\nu_j = -\frac{\omega_{jI}}{K^2} = -\frac{\Lambda_{jR}}{K^2} \label{eq14}
\end{equation} 

\noindent The rationale behind such a definition stems from the behavior of the solution of the multi-dimensional linear
advection-diffusion equation $\frac{\partial f}{\partial t} + {\bf c} \cdot \nabla f = \nu \nabla^2 f$, for which the complex
dispersion relation is given as $\omega = \beta - i\nu K^2$ and the solution of the corresponding Cauchy problem decays with time
as $e^{-\nu K^2 t}$.

One notes that for the vortical mode $V_{gx}^{(1,2)} = c_x$, $V_{gy}^{(1,2)} = c_y$, $V_{gz}^{(1,2)} = c_z$ and $\nu_{(1,2)} = 1/Re$. Hence, the vortical modes are non-dispersive but diffusive in nature and the diffusion coefficient does not depend on the wavenumber. As $\Lambda_3$ is also a real root of Eq.~(\ref{eq12a}), the entropic mode is also non-dispersive and diffusive in nature, like the vortical modes. Therefore, $V_{gx}^3 = c_x$, $V_{gy}^3 = c_y$ and $V_{gz}^3 = c_z$. However, unlike the vortical mode, the diffusion coefficient $\nu_3$ for the entropic mode depends on the wavenumber as well as all the other parameters, \textit{viz}., flow Mach number $M$, the Reynolds number $Re$, bulk viscosity ratio $\kappa/\mu$, ratio of the specific heats $\gamma$ and the Prandtl number $Pr$. The acoustic modes are also diffusive in nature, while these can be dispersive or non-dispersive depending upon the wavenumber. This aspect is illustrated next.

\subsection{Estimation of the bifurcation wavenumber} \label{Bifur}


From Eq.~(\ref{eq12a}) which is a cubic equation with real coefficients, one notes that two scenarios may
arise depending upon whether (a) it has one real and two complex conjugate roots or (b) three real roots. For case-(a),
$\Lambda_{4R} = \Lambda_{5R}$ (and hence, $\nu_4=\nu_5$) and $\Lambda_{4I} = -\Lambda_{5I}$. For this case, the two acoustic
modes have identical diffusion coefficient while these propagate in opposite directions relative to the mean flow. For case-(b),
$\Lambda_{4R} \ne \Lambda_{5R}$ in general (indicating $\nu_4 \ne \nu_5$) and $\Lambda_{4I} = \Lambda_{5I}=0$. Hence for case-(b), two
acoustic modes have different diffusion coefficient while these are non-dispersive in nature. The bifurcation wavenumber $K_b$ between
cases-(a) and (b) can be obtained by setting the discriminant $\Delta$ of the cubic dispersion Eq.~(\ref{eq12a}) to zero. If $\Delta>0$, we have case-(a) while for $\Delta<0$, we have case-(b). It may be
noted that for any general cubic equation with real coefficients, $ax^3+bx^2+cx+d=0$, the expression for the discriminant is given as
$\Delta=\left(b^2c^2-4ac^3-4b^3d-27a^2d^2+18abcd\right)$ \cite{cubic}. For Eq.~(\ref{eq12a}), setting associated $\Delta=0$,
one gets the following cubic equation for the depressed bifurcation wavenumber $\eta_b=\left(K_bM/Re\right)$ as

\begin{eqnarray}
  C_9 z_b^3+C_{10} z_b^2 + C_{11} z_b-4=0 \label{eq15}
\end{eqnarray} 

\noindent where, $z_b=\eta_b^2$, $C_9=(C_{8}^2C_{5}^2-4C_{5}^3)$, $C_{10}=(2C_{5}C_{8}^2-12C_{5}^2-4C_{8}^3C_{7}+18C_{5}C_{7}C_{8})$, $C_{11}=(C_{8}^2-12C_{5}-27C_{7}^2+18C_{7}C_{8})$. The physical value of $\eta_b$ is obtained from the real positive root of
Eq.~(\ref{eq15}). It is to be noted that the coefficients $C_9$, $C_{10}$ and $C_{11}$ are polynomial functions of the parameters
$g=(2-\alpha)=(4/3+\kappa/\mu)$, $\xi_1=\gamma/Pr$ and $\xi_2=1/Pr$. In fact, these coefficients are given as $C_9=g^2\xi_1^2(g-\xi_1)^2$, $C_{10}=2g\xi_1(g+\xi_1)^2-12g^2\xi_1^2-4\xi_2(g+\xi_1)^3+18g\xi_1\xi_2(g+\xi_1)$ and $C_{11}=(g+\xi_1)^2+18\xi_2(g+\xi_1)-27\xi_2^5-12g\xi_1$.      

\section{Results and Discussion} \label{Sec_ii}

\begin{figure} 
\begin{center}
\includegraphics[width=1.0\textwidth]{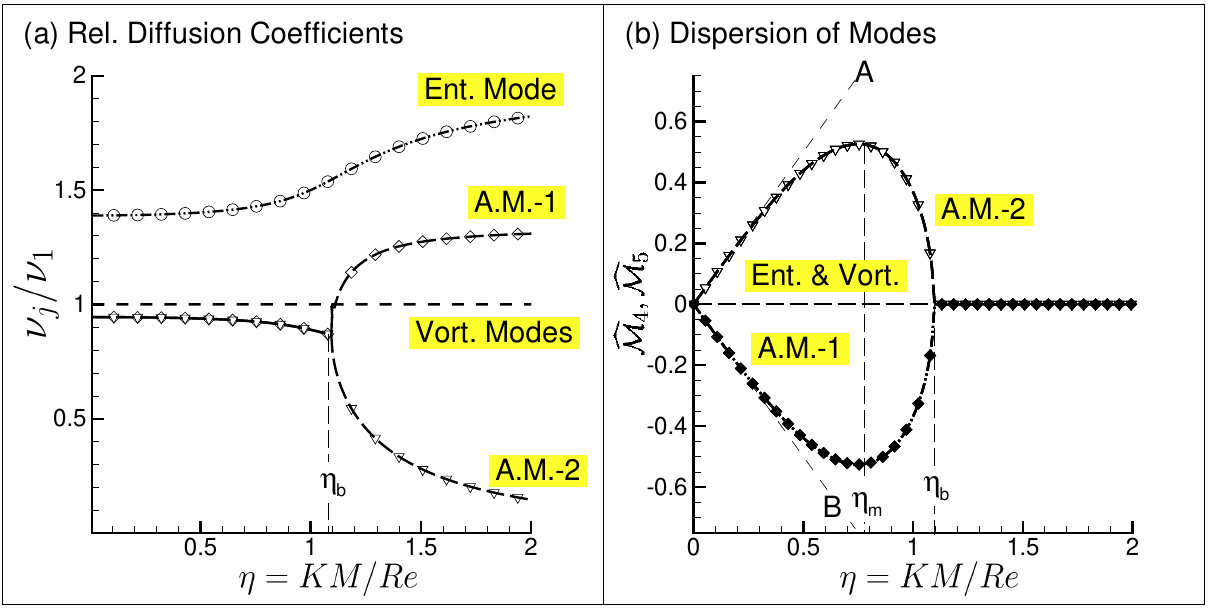}
\caption{(a) $\nu_j/\nu_1$ and (b) $\widehat{\mathcal{M}}_{4,5}$ plotted as a function of the depressed absolute wavenumber $\eta=KM/Re$
  for $\kappa/\mu=0$, $\gamma=1.4$ and $Pr=0.72$. The two straight-lines $A$ and $B$ represents $\widehat{\mathcal{M}}_{4,5}=\mp\eta$,
  respectively.}
\label{fig1}
\end{center}
\end{figure}

Here, we illustrate the dispersive and the diffusive characteristics of the modes. Firstly, we define relative diffusion
coefficient $\nu_j/\nu_1=\lambda_{jR}/\lambda_{1R}$ and dispersion function $\widehat{\mathcal{M}}_j=\lambda_{jI}\eta^2=\Lambda_{jI}\left(M^2/Re\right)$. We illustrate results by mostly plotting these variables as a function of the depressed absolute wavenumber
$\eta=\left(KM/Re\right)$. We chose $\eta$, instead of the absolute wavenumber $K$ as the independent variable, as it makes the 
variables $\nu_j/\nu_1$ and $\widehat{\mathcal{M}}_j$ independent of the Mach number $M$ and the Reynolds number $Re$ of the flow.
These variables when expressed as a function of $\eta$ only depends upon the parameters $\alpha$ (or $\kappa/\mu$), $\gamma$ and the
Prandtl number $Pr$ of the flow. For the inviscid and adiabatic flows (which are governed by the LEE), $\Lambda_{(1,2,3)}=0$ and $\Lambda_{(4,5)}=\mp i \left(K/M\right)$. Therefore, $\widehat{\mathcal{M}}_{(1,2,3)}=0$ and $\widehat{\mathcal{M}}_{(4,5)}=\mp\eta$ for such cases.


\subsection{General dispersive and the diffusive characteristics of the modes}

In Fig.~\ref{fig1}(a) and \ref{fig1}(b), we plot $\nu_j/\nu_1$ and $\widehat{\mathcal{M}}_{4,5}$ as a function of $\eta$ for
$\kappa/\mu=0$, $\gamma=1.4$ and $Pr=0.72$, respectively. The values corresponding to $\gamma$ and $Pr$ correspond to the values for a standard atmosphere while $\kappa/\mu=0$ following the Stokes' hypothesis\cite{Kundu}. Figure~\ref{fig1}(a) shows that the entropic mode is more
diffusive than the vortical mode and the two acoustic modes. The acoustic modes are less diffusive than the vortical mode up to
$\eta=\eta_b$. The depressed bifurcation wavenumber $\eta_b$ corresponds to the situation when the discriminant of the dispersion relation Eq.~(\ref{eq12ab}) becomes zero as noted in Sec.~\ref{Bifur}. For $\eta < \eta_b$ both the acoustic modes have identical diffusion while these display opposite dispersion characteristics. At $\eta=\eta_b$, a bifurcation in the diffusion coefficient for acoustic modes occurs as noted in Fig.~\ref{fig1}(a). For $\eta > \eta_b$, while the diffusion coefficient for the acoustic mode-$1$ increases and is more
than the vortical mode at high-wavenumbers, that for the acoustic mode-$2$ decreases. Figure~\ref{fig1}(b) shows that
$\widehat{\mathcal{M}}_{(1,2,3)}=0$, indicating that entropic and the vortical modes are non-dispersive in nature. For the
acoustic modes $\widehat{\mathcal{M}}_{4}=-\widehat{\mathcal{M}}_{5}$ for $\eta<\eta_b$, indicating that acoustic modes-$1$ and $2$
propagates along opposite directions relative to the mean flow. The two straight-lines $A$ and $B$ in Fig.~\ref{fig1}(b) represents
$\widehat{\mathcal{M}}=\pm\eta$, respectively which indicates the dispersion relation for the upstream and the downstream (relative to the mean flow) propagating acoustic modes for the inviscid and adiabatic case. For $\eta < \eta_c\approx 0.4$, we note $|\widehat{\mathcal{M}}_{(4,5)}|\approx \eta$. Therefore, dispersion (not diffusion) of large wavelength disturbances can still be estimated by LEE. This aspect is illustrated later in Sec.~\ref{ResD_B}. The dispersion function $\widehat{\mathcal{M}}_{(4,5)}$ is dominated by viscous actions for
$0.4 \lesssim \eta \le \eta_b$. At $\eta=\eta_m\approx 0.7\eta_b$ a local minima or maxima is noted to form for the
dispersion function $\widehat{\mathcal{M}}_{4,5}$ corresponding to the acoustic mode-$1$ and -$2$, respectively.  Beyond
$\eta > \eta_b$, $\widehat{\mathcal{M}}_{(4,5)}=0$, indicating that the acoustic modes are non-dispersive in nature.


Here, we may draw the parallel of the dispersion relationship with the Grad’s moment system derived by projecting the Boltzmann equation using physically realizable distribution function. The obtained equations describe the macroscopic governing equations for locally conserved density, momentum and energy, the non-equilibrium stress tensor and energy flux vector etc. and is considered a benchmark for the subsequent theoretical developments in the area of non-equilibrium
thermodynamics \cite{karlin2018}. The dispersion relation of the 1D linearized Grad’s moment system is considered in Refs.~\cite{frouzakis2004} and \cite{struchtrup2003}. The obtained dispersion relation bears a strong resemblance with the current one derived from linearized compressible NSE.  


\begin{figure}[htbp!]
\begin{center}
\includegraphics[width=1.0\textwidth]{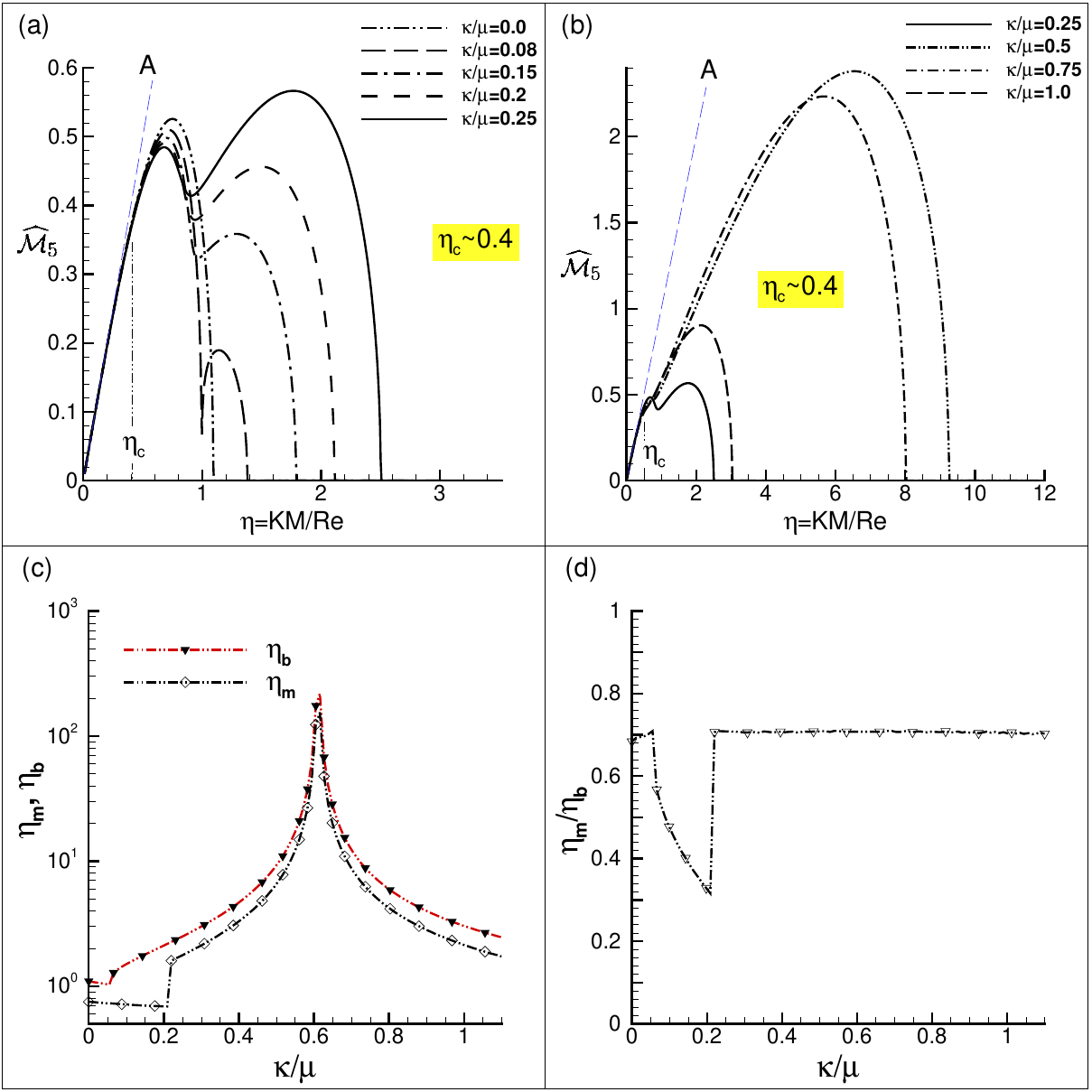}
\caption{(a,b) The dispersion function $\widehat{\mathcal{M}}_5$ for acoustic mode-$2$ plotted as a function of $\eta=KM/Re$ for
  indicated values of $\kappa/\mu$ when $\gamma=1.4$ and $Pr=0.72$. (c,d) $\eta_b$, $\eta_m$ and $\eta_b/\eta_m$ plotted as a
  function of $\kappa/\mu$, respectively.}
\label{fig2}
\end{center}
\end{figure}

\begin{figure}[htbp!]
\begin{center}
\includegraphics[width=0.9\textwidth]{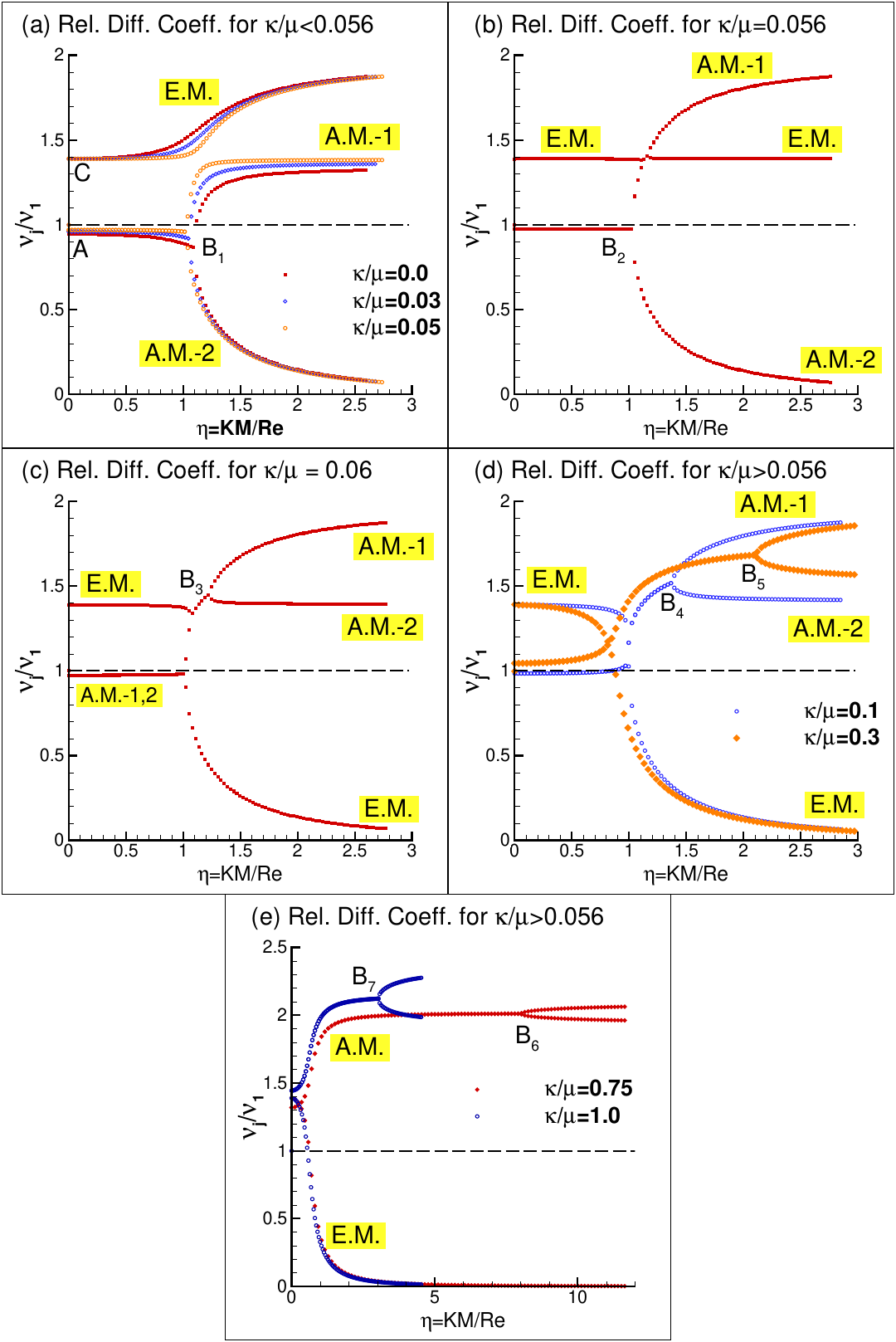}
\vspace{-0.5cm} 
\caption{$\nu_j/\nu_1$ for entropic and acoustic modes plotted as a function of $\eta=KM/Re$ for indicated values of $\kappa/\mu$
  when $\gamma=1.4$ and $Pr=0.72$. The bifurcation points are marked with $B_k$ where $k=1,\cdots, 7$.}
\label{fig3}
\end{center}
\end{figure}

\subsection{Asymptotic solution of the dispersion equation for $\eta<1$}
\label{ResD_B}

When $\eta<1$, generally $\eta<\eta_b$ and the discriminant of Eq.~(\ref{eq12a}) $\Delta>0$. Therefore, the acoustic modes are dispersive in nature indicating $\lambda_3$ is purely real and $\lambda_{4,5}$ are complex conjugates. For the cubic equation given by Eq.~(\ref{eq12ab}) the analytical solution is illustrated in Appendix-$I$. For the case with $\Delta>0$, the solutions for $\lambda_{3,4,5}$ are given as in the form shown through Eqs.~(\ref{Aeq6}). The expression for the discriminant is given in Eq.~(\ref{Aeq4}). Also see Eq.~(\ref{Aeq11}) where $\Delta$ is expressed in terms of $\eta$. From these equations, one can express $\lambda_{3,4,5}$ as 
\begin{eqnarray}
  \lambda_3 & = & \frac{1}{3}\left( \frac{4}{3}+\frac{\kappa}{\mu}+\frac{\gamma}{Pr}\right) + u_1 \\
  \lambda_{4,5} & = & \left[ \frac{1}{3}\left( \frac{4}{3}+\frac{\kappa}{\mu}+\frac{\gamma}{Pr}\right) - \frac{1}{2} u_1 \right] \pm i \frac{\sqrt{3}}{2} v_1
  \label{eq23a} 
\end{eqnarray}

\noindent The expressions for $u_1$ and $v_1$ are given in Sec.~\ref{App_B}. For $\eta<<1$, one can expand $\sqrt{\Delta}$ and subsequently $\sqrt[3]{R_1}$ and $\sqrt[3]{R_2}$ in terms of $\eta$ and thereby obtains the following expressions for $u_1$ and $v_1$ as
$ u_1 = \left( e_1 + e_3 \eta^2 + e_5 \eta^4 + \cdots \right)$ and $v_1 = \left[2/(\sqrt{3}\eta)\right] \left( 1 + e_2 \eta^2 + e_4\eta^4 + e_6 \eta^6 + \cdots \right) $. Putting these in Eq.~(\ref{eq23a}) and noting $z_1/\eta=K/M$, one obtains the dispersion relation for the entropic and the acoustic modes when $\eta<1$ as

\begin{eqnarray}
  \omega_3 & = & \beta - i z_1 \left[ \frac{1}{3}\left( \frac{4}{3}+\frac{\kappa}{\mu}+\frac{\gamma}{Pr}\right) + \left( e_1 + e_3 \eta^2 + e_5 \eta^4 + \cdots \right) \right] \\
  \omega_{4,5} & = & \left\{ \beta \mp \frac{K}{M} \left( 1 + e_2 \eta^2 + e_4\eta^4 + e_6 \eta^6 + \cdots \right) \right\} \nonumber \\
  & & - i z_1 \left[ \frac{1}{3}\left( \frac{4}{3}+\frac{\kappa}{\mu}+\frac{\gamma}{Pr}\right) - \frac{1}{2} \left( e_1 + e_3 \eta^2 + e_5 \eta^4 + \cdots \right) \right]
  \label{eq28a}
\end{eqnarray}

\noindent Therefore, the entropic mode is non-dispersive in nature, whereas for the acoustic modes one gets back the dispersion relation $\omega_{4,5}=\beta \mp K/M$ for inviscid and adiabatic case. One also notes that for $\eta<1$, the relative diffusion coefficients for the entropic and acoustic modes from Eq.~(\ref{eq28a}) are given as

\begin{eqnarray}
  \frac{\nu_3}{\nu_1} & = & \left[ \frac{1}{3}\left( \frac{4}{3}+\frac{\kappa}{\mu}+\frac{\gamma}{Pr}\right) + \left( e_1 + e_3 \eta^2 + e_5 \eta^4 + \cdots \right) \right] \\
  \frac{\nu_{4,5}}{\nu_1} & = & \left[ \frac{1}{3}\left( \frac{4}{3}+\frac{\kappa}{\mu}+\frac{\gamma}{Pr}\right) - \frac{1}{2} \left( e_1 + e_3 \eta^2 + e_5 \eta^4 + \cdots \right) \right]
  \label{eq29a}
\end{eqnarray}

\noindent The coefficients $e_1, \cdots, e_5$ are functions of $b_0$, $b_1$, $d_0, \cdots, d_3$. The last coefficients are given in Sec.~\ref{App_C} in terms of $\kappa/\mu$, $\gamma/Pr$ and $1/Pr$. The coefficients $e_1, \cdots, e_5$ as functions of $b_0$, $b_1$, $d_0 \cdots d_3$ are given as

$ \begin{aligned}
  & e_1 = \left(\frac{2}{27} b_1 \right); \;\;\;\; e_2 = \left( \frac{9}{2} d_2 - \frac{1}{243} b_1^2 \right);  \;\;\;\;
  e_3 = {2} \left( \frac{729}{19683} b_0 - \frac{6561}{19683} b_1 d_2 + \frac{5}{19683} b_1^3 \right); \\
  & e_4 = \left( \frac{9}{2} d_1  - \frac{2}{243} b_0 b_1 + \frac{5}{54} b_1^2 d_2
  - \frac{10}{177147} b_1^4 - \frac{405}{8} d_2^2 \right); \\  
  & e_5 = -2\left( \frac{1}{3} b_0 d_2 + \frac{1}{3} b_1 d_1 - \frac{5}{6561} b_0 b_1^2 - 6 b_1 d_2^2
  + \frac{20}{2187} b_1^3 d_2 - \frac{22}{4782969} b_1^5 \right); \\
  & e_6 = \biggl( \frac{9}{2}d_0 - \frac{405}{4}d_1d_2 - \frac{40}{177147}b_0b_1^3 + \frac{5}{54}b_1^2d_1
  + \frac{55}{19683}b_1^4d_2 - \frac{1}{243}b_0^2 - \frac{154}{129140163}b_1^6 \\
  &     + \frac{13365}{16}d_2^3 - \frac{55}{24}b_1^2d_2^2 + \frac{5}{27}b_0b_1d_2 \biggr) \\
\end{aligned} $

\noindent Therefore, in the long wavelength limit ($\eta<<1$), the relative diffusion coefficients of the entropic and acoustic modes increase linearly with $\kappa/\mu$ and $\gamma$ while these variations are inversely proportional to $Pr$.

\subsection{Effect of bulk viscocity ratio $\kappa/\mu$ on dispersion characteristics} \label{blkdisp}

There are several recent efforts in determining the bulk viscosity of the fluid and its effects on the flow under various conditions.
In \cite{sharma2023}, Green–Kubo method is used to study the bulk viscosity of several dilute gases and their mixtures. Similar efforts are also performed in \cite{okumura2002},\cite{cramer2012},\cite{jaeger2018} to determine $\kappa$ for various gases using molecular dynamics simulations. The bulk viscosity ratio $\kappa/\mu$ have been estimated to range from a very small number to $4$ for Argon \cite{jaeger2018}, $0.01-2$ for gaseous \ch{CO2}\cite{jaeger2018}, from $0.5-0.7$ for \ch{N2} \cite{sharma2023}, from $0.5-0.6$ for \ch{O2} and similar values for \ch{N2}-\ch{O2} mixture\cite{sharma2023} and from $0.6-1.75$ for dry air \cite{sharma2023,cramer2012,shang2019} depending upon density and temperature.

In Figs.~\ref{fig2}(a,b), we plot the dispersion function $\widehat{\mathcal{M}}_{5}$ as a function of $\eta$ for indicated $\kappa/\mu$
cases with $\gamma=1.4$ and $Pr=0.72$. As noted in Fig.~\ref{fig1}(b), $\widehat{\mathcal{M}}_{5}$ for $\kappa/\mu=0$, initially
monotonically increases, reaches maxima at $\eta=\eta_m$ and subsequently decreases to zero at $\eta=\eta_b$. With increase in
$\kappa/\mu$ from $0$ to $0.25$, a secondary peak is noted to gradually arise. For all these cases, however,
$\widehat{\mathcal{M}}_{5}\approx \eta$ for $\eta \lesssim 0.4$. This indicates that change in bulk viscosity in this regime has less significant 
effect on the dispersion function for the spectral components having wavenumbers smaller than $\eta \lesssim 0.4$, and these follow the inviscid isentropic dispersion function given by LEE. This may be noted by comparing the deviation of $\widehat{\mathcal{M}}_{5}$ with the line $A$ which depicts the variation of the dispersion function from LEE. A detailed analysis in the low wavenumber regige is provided in the previous subsection.     

The increase in $\kappa/\mu$ from $0$ to $0.25$ is noted to significantly alter the dispersion relation of higher wavenumber components firstly due to the appearance of a secondary peak, whose extent is noted to grow from $\kappa/\mu=0.08$ to $0.25$ and secondly due to increase in $\eta_b$. For $\kappa/\mu=0.25$, we note the secondary peak to be more dominant than the primary one. As the bulk viscosity ratio is further increased from $0.25$ we note the initial primary one to merge with the secondary one and thereby looses its distinct signature. The previous secondary peak become the sole dominant one for $\kappa/\mu \gtrsim 0.3$. One also notes from Fig.~\ref{fig2}(b) that $\eta_b$ to first increase, reaches a maximum value and subsequently decrease with increase in $\kappa/\mu$.

The last aspect is further explored in Fig.~\ref{fig2}(c) where, $\eta_b$ and $\eta_m$ is plotted as a function of $\kappa/\mu$. One notes from this figure that $\eta_b$ to be maximum at $\kappa/\mu \simeq 0.6$. The corresponding value of $\eta_b$ at $\kappa/\mu \approx 0.6$ is roughly two orders of magnitude higher than that at $\kappa/\mu=0$. The value of $\eta_m$ corresponding to the peak location for the dispersion function $\widehat{\mathcal{M}}_{5}$ remains virtually constant up to $\kappa/\mu \simeq 0.2$, at which point a discontinuous change is noted. This is related with almost unchanged location of the primary peak, appearance of a secondary peak and the eventual dominance of the latter beyond $\kappa/\mu \gtrsim 0.2$. It may be noted that beyond $\kappa/\mu \simeq 0.2$, the $\eta_m$ follows similar trend as that shown by $\eta_b$. In fact, for $\kappa/\mu \gtrsim 0.2$, $\eta_m/\eta_b$ remains almost constant at $0.7$ as shown in Fig.~\ref{fig2}(d).

\subsection{Effect of bulk viscocity ratio $\kappa/\mu$ on diffusion characteristics}
\label{effdiffbulk}


Next, we illustrate the variation of the relative diffusion coefficient $\nu_j/\nu_1$ for the entropic and the acoustic modes
when $\kappa/\mu$ is varied for $\gamma=1.4$ and $Pr=0.72$. This is shown in Fig.~\ref{fig3} where the abbreviations $E.M.$, 
$A.M.-1$ and $A.M.-2$ denote the entropic, acoustic modes-$1$ and -$2$, respectively. We note three different trends depending
upon whether $\kappa/\mu$ is less than, equal to or more than $0.56$. The first trend is noted
when $\kappa/\mu<0.56$. some typical cases are shown in Fig.~\ref{fig3}(a). For this case, following observations may be noted:

\begin{enumerate} 
\item The entropic mode is more diffusive than the vortical mode while $\nu_3/\nu_1$ increases with increase in $\eta$ as the coefficient $e_3$ in Eq.~(\ref{eq29a}) is positive.  
\item The acoustic modes are less diffusive and $\nu_{(4,5)}/\nu_1$ is noted to decrease with increase in $\eta$ for $\eta<\eta_b$ as $e_3>0$. The bifurcation point $\eta_b$ is marked as $B_1$ in Fig.~\ref{fig3}(b).
\item For $\eta>\eta_b$, while the relative diffusion coefficient of the acoustic mode-$1$ is intermediate between the vortical and the entropic modes, it monotonically decrease for the acoustic mode-$2$ with increase in $\eta$.
\end{enumerate}

In all subsequent discussions, we denote similar trend for the variation of the relative diffusion coefficients as $T_1$ for easy reference. For the cases shown in Fig.~\ref{fig3}(a), we also note that with increase in $\kappa/\mu$, $\nu_3/\nu_1$ decreases and tends to become more flatter. Similarly, the level of $\nu_5/\nu_1$ increases with increase in $\kappa/\mu$ beyond $\eta=\eta_b$ and likewise tend to become flatter for high $\eta$ components.

At $\kappa/\mu=0.56$, the $\nu_3/\nu_1$ becomes independent of $\eta$, which is formed by connecting the increasingly flatter
parts of the entropic and the acoustic mode-$2$ beyond the bifurcation wavenumber as noted in Fig.~\ref{fig3}(b). The relative diffusion
coefficients of the acoustic modes are also noted to be independent of $\eta$ up to the bifurcation point $B_2$. Such a variation is noted probably because all the coefficients $e_3$, $e_5,\cdots$ bacome vanishingly small for this combination of parameters. Such a trend of variation is denoted hereafter as $T_0$ for easy reference. Trend $T_0$ indicates that the entropic and acousic modes are essentially decoupled and the entropic mode effectively follows a convection-diffusion equation. 

Beyond $\kappa/\mu>0.56$, altered variation of the relative diffusion coefficients are noted which become more prominant as $\kappa/\mu$ increases further. For these cases, while $\nu_3/\nu_1$ monotonically decreases, the $\nu_{(4,5)}/\nu_1$ for $\eta<\eta_b$ monotonically increases with $\eta$. Not only these variation become steeper but also a gradual increase in the level of relative diffusion coefficients for the acoustic modes before the bifurcation point occurs with increase in $\kappa/\mu$. Similar trends are subsequently referred to as $T_2$. We later see these trends are generic and also noted when $\gamma$ or $Pr$ is parametrically varied.

\begin{figure}[htbp!]
\begin{center}
\includegraphics[width=1.0\textwidth]{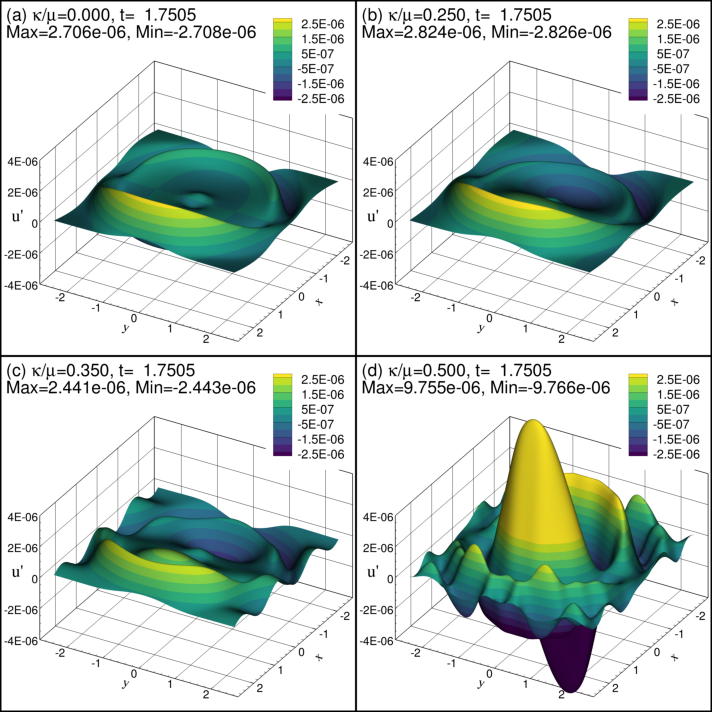}
\caption{Disturbance $u$-velocity plotted in the $(x,y)$-plane at $t=1.75$ for indicated values of $\kappa/\mu$.
  Here, $M=0.8$, $Re=100$, $\gamma=1.4$ and $Pr=0.72$.}
\label{fig4}
\end{center}
\end{figure}

\begin{figure}[htbp!]
\begin{center}
\includegraphics[width=1.0\textwidth]{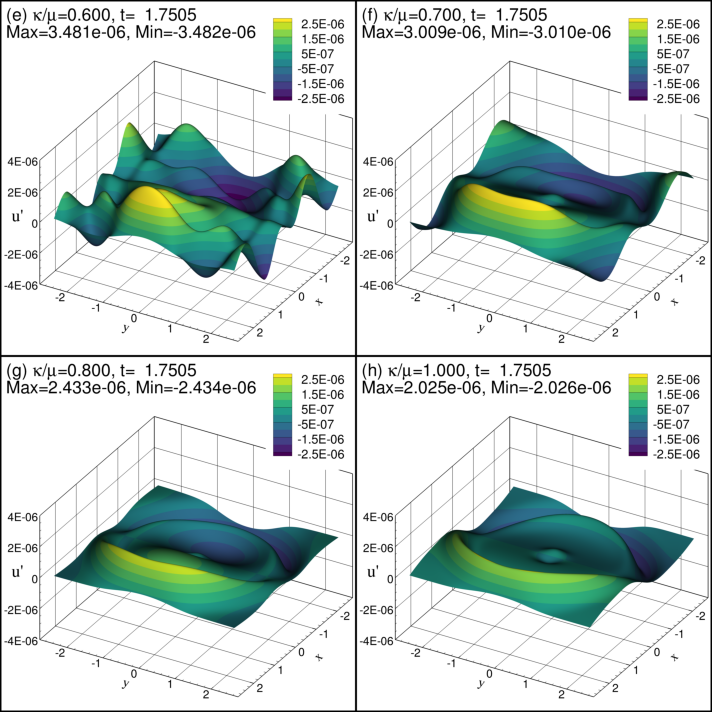}
\caption{Disturbance $u$-velocity plotted in the $(x,y)$-plane at $t=1.75$ for indicated values of $\kappa/\mu$.
  Here, $M=0.8$, $Re=100$, $\gamma=1.4$ and $Pr=0.72$.}
\label{fig5}
\end{center}
\end{figure}

\subsection{Linearized disturbance evolution}
\label{linearized} 

To asses the implication of the above analysis on the effects of bulk viscosity ratio, we perform the linearized evolution of
disturbances by varying bulk viscosity ratio $\kappa/\mu$. We specifically perform the initial value problem, where the initial
condition in the form of a wave-packet containing multiple spectral components is imposed as a combination of entropic and acoustic disturbances. Solution of similar test problem was also reported in \cite{pirozzoli2013}. The exact solution of the linearized $NSE$ given by Eqs.~(\ref{eq7}-\ref{eq8}) may be obtained by considering the Fourier transform as given by Eq.~(\ref{eq10}). The evolution equation of the Fourier amplitudes are given as  
\begin{eqnarray}
\frac{\partial }{\partial t} \left\{ \mathcal{F} \right\}+ [D] \left\{ \mathcal{F} \right\}=0
\label{eq50} 
\end{eqnarray} 

\noindent where, $\left\{ \mathcal{F} \right\} = \left( \xi, \hat{\bf \Phi}, \theta \right)^T$ and the 
matrix $[D]$ indicates the corresponding evolution matrix. From Eq.~(\ref{eq50}), one obtains the exact solution for 
$\mathcal{F} (t)$ considering a periodic problem as 
\begin{eqnarray}
\left\{ \mathcal{F} \right\} = [R] \; e^{\left\{-[\chi] t\right\}} \; [L] \; \left\{ \mathcal{F}_0 \right\}
\label{eq52} 
\end{eqnarray}     

\noindent where $\left\{ \mathcal{F}_0 \right\}$ is the Fourier amplitudes corresponding to the initial perturbations, while 
$[R]$, $[L]$ and $[\chi]$ are the right-eigenvector, left-eigenvector and eigenvalue matrices corresponding to $[D]$ such that $[D]=[R] [\chi] [L]$. Therefor, diagonal entries of $[\chi]$ are given as $\chi_j = i\omega_j=i\beta-\Lambda_j$, so that $\Lambda_j$ are the roots of the dispersion equation given by Eq.~(\ref{eq12}). The right-eigenvector appearing in Eq.~(\ref{eq50}) for a $3D$-problem may be symbolically given as 
\begin{eqnarray}
[R] = \begin{bmatrix}
 r_1 & r_2 & r_j   
\end{bmatrix} = 
\begin{bmatrix}
 0                             & 0                              & 1  \\
{\bf K}\times {\bf{\delta}}_1/K & {\bf K}\times {\bf{\delta}}_2/K & -i{\bf K} F_j/{z_1 M^2} \\
0                              & 0                              & \left(\gamma-1\right)H_j 
\end{bmatrix}
\label{eq51} 
\end{eqnarray}   

\noindent where $j=3,4,5$; ${\bf{\delta}}_1=(0,0,1)^T$; ${\bf{\delta}}_2=(0,1,0)^T$; $F_j = \gamma \left( \lambda_j + 1/Pr \right)/\left[ \left( \lambda_j + (2-\alpha)\right)\left(\lambda_j+\gamma/Pr \right)\right]$; $H_j=\lambda_j/\left(\lambda_j+\gamma/Pr \right)$ and $\lambda_j$ are the roots of the characteristic Eq.~(\ref{eq12ab}) for entropic and acoustic modes. Here, the column vectors $r_1$, $r_2$, $r_3$, $r_4$ and $r_5$ represents eigenvectors 
corresponding to two vortical, entropic, acoustic mode-$1$ and -$2$, respectively. One notes that while the perturbation velocity components corresponding to the vortical eigenvectors are solenoidal, those corresponding to entropic and acoustic modes
are irrotational. This follows the theoretical criterion postulated in \cite{doak1989} to identify the entropic and acoustic modes. 

For the ease of computation, here we consider only the a $2D$-problem. It may be noted that $2D$ compressible NSE supports four waves, one vortical, one entropic and two acoustic waves. The dispersion relation of the entropic and two acoustic waves for a $2D$ problem is still given by Eq.~(\ref{eq12a}) or Eq.~(\ref{eq12ab}), only difference being that here the wavenumber vector ${\bf K}=\left(k_x,k_y\right)$ and therefore, $K=\sqrt{k_x^2+k_y^2}$. The right eigenvector described in Eq.~(\ref{eq51}) is slightly modified for the $2D$-problem which supports only one vortical mode. The spectral nature of entropic and acoustic modes with respect to the absolute wavenumber remain identical. Once $\mathcal{F}(t)$ is obtained following 
Eq.~(\ref{eq52}), the original perturbation quantities can be computed by performing inverse Fourier transform. In actual computation, we first perform fast Fourier transform (FFT) of the initial perturbation, time advance $\mathcal{F}$ using Eq.~(\ref{eq52}) and obtain back the exact solution at the intended time instant by performing inverse FFT (IFFT).

The physical domain extends from $-2.5$ to $2.5$ along each directions and we used a total of $501$ uniformly distributed points
to represent the physical domain with a grid spacing of $\Delta x=\Delta y=0.01$. The results are reported for $M=0.8$, $Re=100$, $\gamma=1.4$ and $Pr=0.72$ while the mean flow is directed at an angle of $\theta=\tan^{-1}\left(c_y/c_x\right)=45^{\circ}$. We impose the initial disturbance as  

\begin{eqnarray} 
p'(x,y,0) & = & p_a(x,y)   \nonumber \\
\rho'(x,y,0) & = & \frac{1}{\gamma}p_a(x,y)  - \theta_e(x,y) \nonumber \\
T'(x,y,0) & = & \left(\frac{\gamma-1}{\gamma}\right) p_a(x,y) + \theta_e(x,y) \nonumber \\
p_a(x,y) = \theta_e(x,y) & = & \left( \sum_{m,n} \cos\left[k_{0x_m} \left(x-x_0\right)+\varphi_{xm} \right]\;\cos\left[k_{0y_n} \left(y-y_0\right)+\varphi_{yn}\right] \right)\; e^{\left(-\sigma r^2\right)} \nonumber \\
u'(x,y,0)& = & v'(x,y,0)=0
\label{acp}    
\end{eqnarray} 

\noindent where $r=\sqrt{\left(x-x_0\right)^2+\left(y-y_0\right)^2}$ while $p_a$ and $\theta_e$ indicate initial disturbances corresponding to acoustic and the entropic pulse, respectively. The center of the pulse is initially located at the position
$(x_0,y_0)=(0.25,0.25)$. Here, $k_{0x_m}$ and $\varphi_{x_m}$ are the central wavenumber of the wave-packet and the phase, respectively along the $x$-direction for the $m^{th}$-mode. We use a total of $m=101$ and $n=101$ discrete wavenumbers along $x$- and $y$-directions, respectively, such that $k_{0x_m}\Delta x$ and $k_{0y_n}\Delta y$ each spans from $0.1$ to $2.0$. For any discrete computation, maximum allowable resolution is limited by the Nyquist limit such that $k_{0x_{max}}\Delta x=k_{0y_{max}}\Delta y=\pi$. The phase $\varphi_{x_m}$ and $\varphi_{y_n}$ is specified such that it varies randomly between $0$ and $2\pi$ with uniform probability distribution. We note that the depressed wavenumber $\eta$ for the initial imposed disturbances spans from $\eta_{min}=0.08$ to $\eta_{max}=1.6$. Here, the initial perturbation entropy $s'(x,y,0)=\theta_e$ and vorticity $\Omega'(x,y,0)=0$.

Figures~\ref{fig4} and \ref{fig5} shows the disturbance $u$-velocity component at $t=1.75$ for indicated values of $\kappa/\mu$.
We have also noted the maximum and minimum values of the
disturbance in these frames for a specific value of $\kappa/\mu$. In Fig.~\ref{fig4}(a), when $\kappa/\mu=0$, one notes two peripheral lobes with a dip at the center. As one increases $\kappa/\mu$, the intensity of the wavy disturbances are noted to increase. At $\kappa/\mu=0.5$, we note significantly higher amount of disturbance as compared to $\kappa/\mu=0$, $0.25$ and $0.3$. Further increase in $\kappa/\mu$ gradually brings the disturbance to the level of $\kappa/\mu=0$, \textit{i.e.}, the case with zero bulk viscosity. The topology of the $u'$-surface for higher $\kappa/\mu$ cases ($\kappa/\mu=0.75$ and $1.0$) are similar to the lower $\kappa/\mu$ cases ($\kappa/\mu=0.0$ and $0.25$).
The results indicate that effects of the bulk viscosity ratio is significant when $\kappa/\mu \simeq 0.5$ where relatively significant disturbance level is noted. For this bulk viscosity ratio the relative diffusion coefficients are less and the extent of dispersive wavenumbers are large. Similar results are also noted when only acoustic disturbances are imposed as the initial condition (results not reported here). We note that according to Fig.~\ref{fig2}(c), $(\eta_b)_{max}$ is noted for $\kappa/\mu\simeq 0.6$ which is close to the bulk viscosity ratio for which relatively significant disturbance evolution is observed here. Here lies the importance of Fig.~\ref{fig2}(c) which specifies the range of $\kappa/\mu$ that would yield significant acoustic and entropic disturbance evolution, radiation and dispersion. This range of $\kappa/\mu$ may be identified by those, that are close to the one for which $\eta_b$ is maximum. Moreover, as dispersion may enhance the process of energy cascade in the presence of nonlinearities \cite{cai2002}, these range of $\kappa/\mu$ is expected to play a crucial role in the generation of fine-scale turbulence along with playing an important role in acoustic radiation. These aspects may be investigated by direct numerical simulation (DNS) of a system with bulk viscosity ratios laying in these range.

\subsection{Physical estimation of the minimum wavelength for the dispersive waves}

It is imperative to give a physical explanation of the depressed wavenumber $\eta=KM/Re$. Expressing $M=U_m/c$ and $Re=\rho_mU_mL/\mu_m$,
where $c=\sqrt{\gamma R T_m}$ is the speed of sound, one obtains $\eta=(2\pi/\lambda^*)(\mu_m/\rho_m c)$. Here, $\lambda^*$ is the dimensional wavelength of the disturbances. For an estimate, consider air at standard atmospheric pressure ($101\;KPa$) and temperature of $20^{\circ}$. At these standard atmospheric conditions, the density $\rho_m$, coefficient of dynamic viscosity $\mu_m$ and speed of sound $c$ are approximately $\rho_m \simeq 1.225 Kg/m^3$, $\mu_m\simeq 1.81\times 10^{-5} Pa.s$ and $c\simeq 330m/s$. Therefore, the ratio $(\mu_m/\rho_m c)$ comes out to be $4.48\times 10^{-8}\;m$. Hence, considering $\eta_b\approx \mathcal{O}(1)$, the smallest wavelength of dispersive acoustic disturbances is roughly $\lambda^*_{min}\approx 2.81\times 10^{-7}\;m$ whereas, at standard atmospheric conditions the mean free path of the air molecules is roughly $l=66\;nm$ \cite{jennings1988}. Therefore, for this case $\lambda^*_{min}$ is one order of magnitude higher than $l$ and high-wavenumber acoustic dispersive disturbances as predicted by NSE are at the boundaries of the continuum hypothesis \cite{white1966,karniadakis2006}. When the bulk viscosity ratio $\kappa/\mu$ is in the range of $0.4-0.6$, $\lambda^*_{min}$ becomes of the same order as $l$. These scales are, however, smaller than Kolmogorov length-scale $\bar{\eta}_k=Re^{-3/4}l_0$ which defines the length-scales associated with locally isotropic smallest eddies in a highly turbulent flow, where $l_0$ is the integral length scale \cite{pope2000}. For a turbulent flow with $Re\simeq 10^{6}$, Kolmogorov length-scale $\bar{\eta}_k\simeq 3.2\times 10^{-5}m$. In Table~\ref{tab1} we tabulate the properties of some common gases like Carbon dioxide (\ch{CO2}), Oxygen (\ch{O2}), Carbon monoxide (\ch{CO}), Hydrogen (\ch{H2}), Nitrogen (\ch{N2}), dry air and Argon (\ch{Ar}) at atmospheric pressure and $T=300K$ and $900K$. We provide a column for $\lambda^*_{min}$ assuming $\eta_b=1$. One notes that $\lambda^{*}_{min}$ is almost $5$ times more than $l$ for most of the gases.      

\begin{table}[]
\centering
\begin{adjustbox}{width=1.0\textwidth}
\small
\begin{tabular}{|c|c|c|c|c|c|c|c|c|c|}  
\hline
Name of & $T$   & $\rho$     & $\gamma$ & $c$     & $\mu$    & $\hat{k}$          & $\Pr$ & $l$    & $\lambda^{*}_{min}  $ \\
the Gas & ($K$) & ($Kg/m^3$) &          & ($m/s$) & ($Pa.s$) & ($Watt/K.m$) &        & ($m$) & ($m$)  \\ \hline
\multirow{2}{*}{\ch{CO2}} & $300$ & $1.796$ & $1.293$ & $268.31$ & $1.49\times10^{-5}$ & $0.0166$ & $0.769$ & $8.45\times10^{-8}$ & $1.95\times10^{-7}$ \\ \cline{2-10} 
                          & $900$ & $0.596$ & $1.186$ & $447.18$ & $3.61\times10^{-5}$ & --       & --      & $2.53\times10^{-7}$ & $8.52\times10^{-7}$ \\ \hline
\multirow{2}{*}{\ch{O2}} & $300$ & $1.301$ & $1.396$ & $329.62$ & $2.06\times10^{-5}$ & $0.0268$ & $0.709$ & $7.69\times10^{-8}$ & $3.02\times10^{-7}$ \\ \cline{2-10} 
                         & $900$ & $0.433$ & $1.319$ & $555.66$ & $4.45\times10^{-5}$ & -- & -- &$2.31\times10^{-7}$ & $1.16\times10^{-6}$ \\ \hline
\multirow{2}{*}{\ch{CO}} & $300$ & $1.138$ & $1.401$ & $353.10$ & $1.79\times10^{-5}$ & $0.0252$ & $0.737$ & $6.51\times10^{-8}$ & $2.79\times10^{-7}$\\ \cline{2-10} 
                         & $900$ & $0.379$ & $1.343$ & $599.06$ & $3.89\times10^{-5}$ & -- & -- & $1.95\times10^{-7}$ & $1.08\times10^{-6}$ \\ \hline
\ch{H2} & $300$ & $0.082$ & $1.405$ & $1319.11$ & $8.96\times10^{-6}$ & $0.1817$ & $0.706$ & $1.1\times10^{-7}$ & $5.21\times10^{-7}$\\ \hline 
\multirow{2}{*}{\ch{N2}} & $300$ & $1.138$ & $1.401$ & $353.13$ & $1.79\times10^{-5}$ & $0.0261$ & $0.714$ & $6.94\times10^{-8}$ & $2.79\times10^{-7}$\\ \cline{2-10} 
                         & $900$ & $0.379$ & $1.35$ & $600.8$ & $3.75\times10^{-5}$ & $0.0603$ & $0.712$ & $2.08\times10^{-7}$ & $1.03\times10^{-6}$ \\ \hline 
\multirow{2}{*}{Dry Air} & $300$ & $1.177$ & $1.4017$ & $347.33$ & $1.85\times10^{-5}$ & $0.0262$ & $0.708$ & $6.66\times10^{-8}$& $2.84\times10^{-7}$ \\ \cline{2-10} 
                         & $900$ & $0.392$ & $1.345$ & $589.65$ & $3.9\times10^{-5}$ & $0.0628$ & $0.696$ & $1.73\times10^{-7}$ & $1.06\times10^{-6}$ \\ \hline  
\multirow{2}{*}{\ch{Ar}} & $300$ & $1.624$ & $1.67$ & $322.66$ & $2.29\times10^{-5}$ & $0.0177$ & $0.677$ & $7.96\times10^{-8}$ & $2.75\times10^{-7}$ \\ \cline{2-10} 
                         & $900$ & $0.541$ & $1.667$ & $559.11$ & $5.06\times10^{-5}$ & $0.0398$ & $0.661$ & $2.39\times10^{-7}$ & $1.05\times10^{-6}$ \\ \hline            
\end{tabular}
\end{adjustbox}
\caption{Properties of some common gases at atmospheric pressure and $T=300K$ and $900K$. All the variables are in $SI$ units. The data for density $\rho$, specific heat ratio $\gamma$, speed of sound $c$, dynamic viscosity $\mu$, thermal conductivity $\hat{k}$ and Prandtl number $Pr$ are taken from \cite{hilsenrath1955}. The mean free path is calculated assuming classical Maxwell-Boltzmann theory as $l=k_BT/\left(\sqrt{2}\pi d^2 P\right)$ \cite{sommerfeld1964}, where $k_B$ is the Boltzmann constant and $d$ is the effective diameter of the molecule. Also, $\lambda^*_{min}=2\pi \mu/\left(\rho c\right)$ is the minimum wavelength of the dispersive acoustic disturbances assuming $\eta_b=1$.}
\label{tab1}
\end{table}



\begin{figure}[htbp!]
\begin{center}
\includegraphics[width=1.0\textwidth]{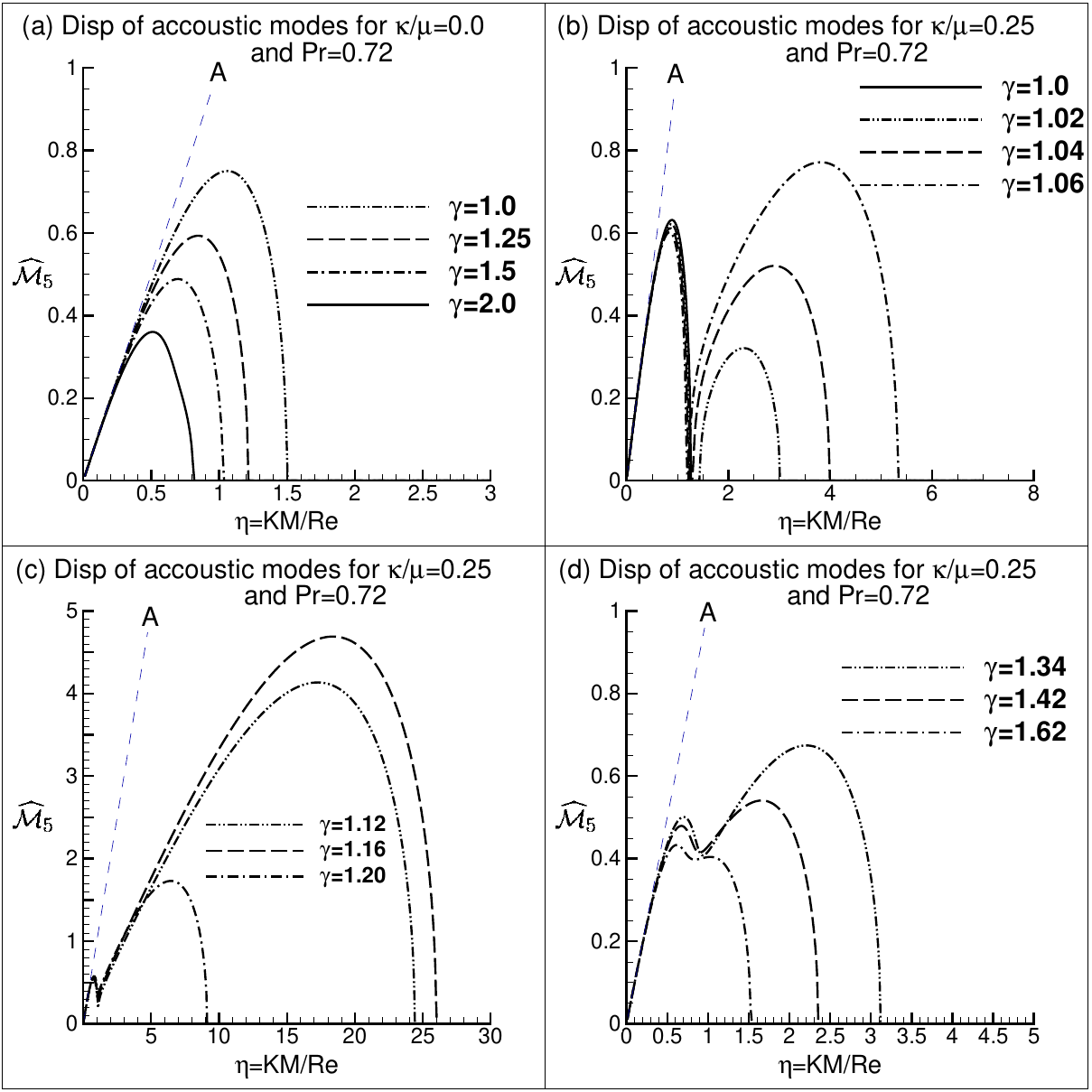}
\caption{The dispersion function $\widehat{\mathcal{M}}_5$ for the acoustic mode-$2$ plotted as a function of $\eta=KM/Re$ for
  indicated values of $\gamma=1.4$ when (a) $\kappa/\mu=0.0$ and (b,c,d) $\kappa/\mu=0.25$. Here, and $Pr=0.72$. The straight-line $A$
  represents $\widehat{\mathcal{M}}_5=\eta$, \textit{i.e.}, the dispersion function for the inviscid and adiabatic case.}
\label{fig6}
\end{center}
\end{figure}

\begin{figure}[htbp!]
\begin{center}
\includegraphics[width=1.0\textwidth]{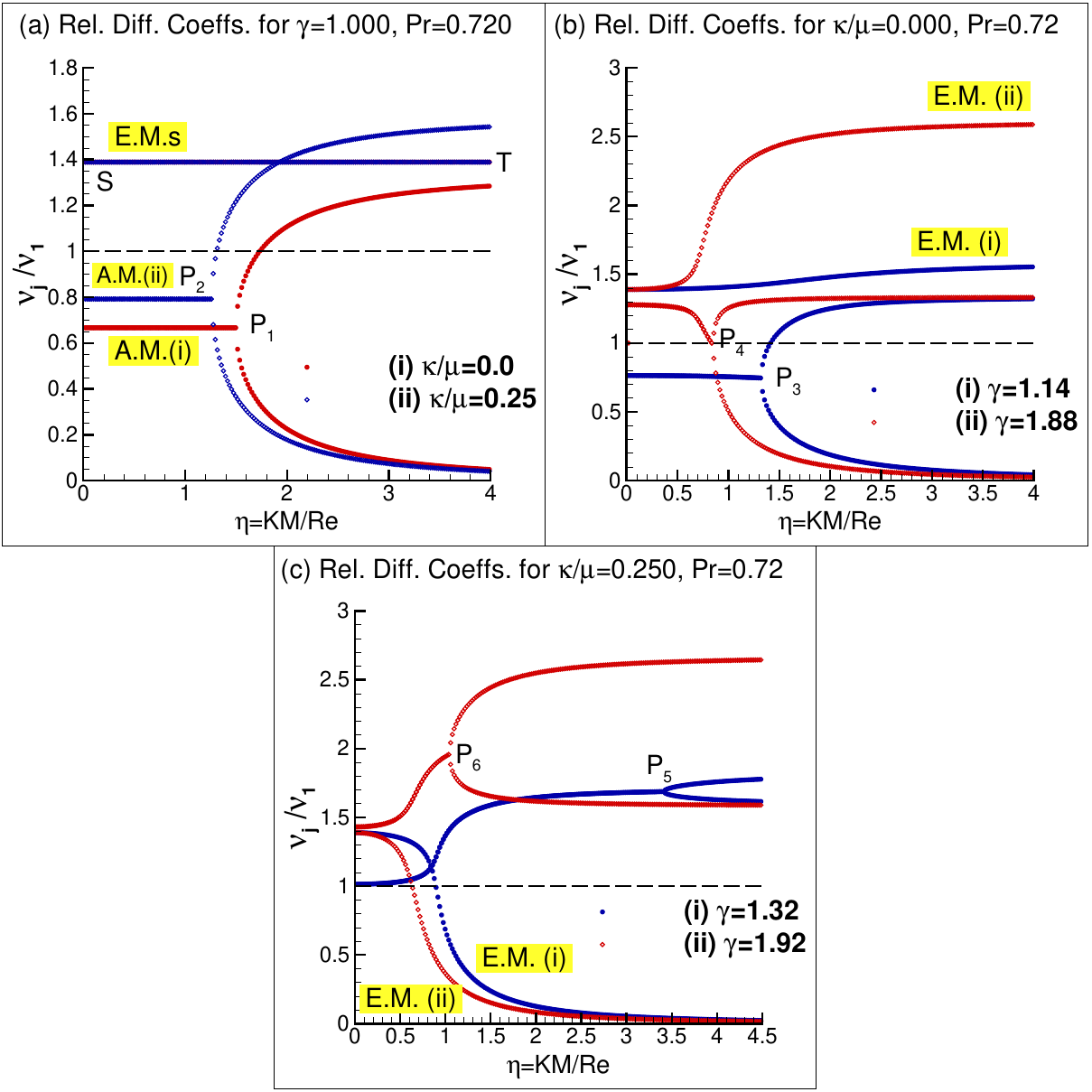}
\caption{$\nu_j/\nu_1$ for entropic and acoustic modes plotted as a function of $\eta=KM/Re$ for indicated values of $\gamma$ and
  $\kappa/\mu$ when $Pr=0.72$. The bifurcation points are marked with $P_k$ where $k=1,\cdots, 5$.}
\label{fig7}
\end{center}
\end{figure}

\begin{figure}[htbp!]
\begin{center}
\includegraphics[width=1.0\textwidth]{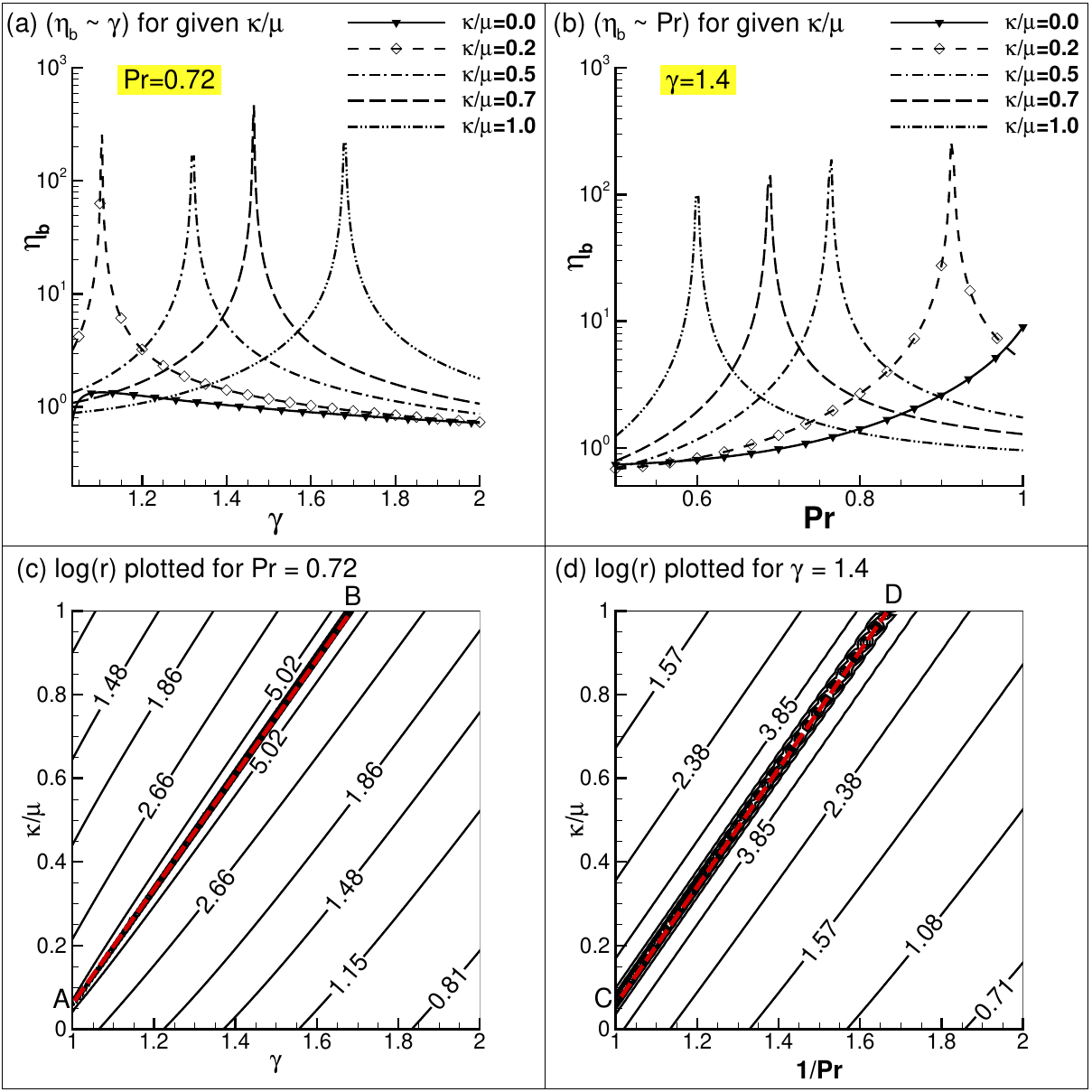}
\caption{(a,b) $\eta_b$ plotted as a function of $\gamma$ and $Pr$ for indicated values of $\kappa/\mu$, respectively.
  (c,d) $\log(\hat{r})$ for the cubic Eq.~(\ref{eq15}) corresponding to $\eta_b^2$ shown plotted in $(\kappa/\mu,\;\gamma)$ and
  $(\kappa/\mu,\; 1/Pr)$-planes for indicated values of $Pr$ and $\gamma$, respectively. Here, $\hat{r}$ is the radius of the circle
  in the complex plane on which the three real solutions of any cubic Eq.~(\ref{eq15}) lie for negative discriminant
  (See Eq.~(\ref{Aeq5}) in Appendix-I).}
\label{fig8}
\end{center}
\end{figure}


\subsection{Effects of specific heat ratio $\gamma$ and Prandtl number $Pr$ on the dispersion relation}


Apart from the bulk viscosity ratio, two other important parameters in the dispersion relation is the specific heat ratio $\gamma$ and the Prandtl number $Pr=c_p\mu/\hat{k}$. As $\gamma=c_p/c_v$, $\gamma\geq 1.0$ following the second law of thermodynamics \cite{Kundu}. For gases with mono-atomic molecules, the classical Maxwell-Boltzmann distribution specifies $\gamma=5/3$, whereas for the gases with diatomic or polyatomic molecules, these values theoretically are given as $7/5$ and $4/3$, respectively considering the molecules as rigid particles connected by massless links \cite{sommerfeld1964}. Similarly, the classical equipartition of energy predicts that $\gamma$ is related to the thermally accessible degrees of freedom ($f$) as $\gamma=1+1/f$ \cite{sommerfeld1964}. For common gases like \ch{CO2}, \ch{O2}, \ch{CO}, \ch{H2}, \ch{N2}, air, Argon etc., the value of $\gamma$ is noted to decrease with temperature \cite{white1966} (only Argon displays very marginal increase at high temperature). For these gases, the value of $\gamma$ varies between $1.41-1.17$, $1.42-1.19$, $1.43-1.28$, $1.65-1.3$, $1.42-1.19$, $1.42-1.19$ and $1.70-1.65$, respectively when temperature is varied from very low to very high\cite{white1966}. Also, refer to Table~\ref{tab1} for typical values of $\gamma$ for some common gases at standard atmospheric pressure and two temperatures of $300K$ and $900K$.         

Here, we investigate the specific heat ratio by varying it from $\gamma=1.0$ to $2.0$. For $\gamma=1.0$, Eq.~(\ref{eq12ab}) may be solved analytically and the corresponding solutions are given as
\begin{eqnarray}
  \lambda_3=-\frac{1}{Pr} \;\; \text{and} \;\; \lambda_{4,5}=-\alpha_*\pm \sqrt{\alpha_*^2-\frac{1}{\eta^2}}
  \label{eq26a} 
\end{eqnarray}
\noindent where $\alpha_*=\left(2/3+\kappa/(2\mu)\right)$. Therefore, $\nu_3/\nu_1=1/Pr$ and is independent of the depressed wavenumber. Similarly, before the bifurcation point, the relative diffusion coefficients of the acoustic modes are also constant at $\nu_{4,5}/\nu_1=\alpha_*$ while $\eta_b=1/\alpha_*$. Following Eq.~(\ref{eq8}), we note  that for $\gamma=1$, the entropic mode follows a convection-diffusion equation (for $\gamma=1$, $s'=T'$). 

In Fig.~\ref{fig6}, we show the dispersion function $\widehat{\mathcal{M}}_5$ for the acoustic modes as a function of $\eta=KM/Re$ for indicated values of $\gamma=1.4$. While the variation corresponding to $\kappa/\mu=0.0$ is shown in frame (a), frames (b,c,d) are for $\kappa/\mu=0.25$. For all the plots Prandtl number is fixed at $0.72$. From Fig.~\ref{fig6}(a), we note that the dispersive zone keeps shrinking with increase in $\gamma$ for $\kappa/\mu=0$, consequently, the range of wavenumber zone for which one can apply the dispersion relation for the inviscid and the adiabatic case (from LEE) also decrease with increase in $\gamma$. For
$\kappa/\mu=0.25$, the trend of the dispersion function is similar to what is noted in Fig.~\ref{fig2}(a,b) when the bulk viscosity ratio was varied. When $\gamma$ is close but slightly more than one, we note the existence of two dispersive zones (see the curve for $\gamma=1.02$ in Fig.~\ref{fig6}(b)). Subsequent increase in $\gamma$ increases $\eta_b$ as well as the extent of the zone where deviation from inviscid and adiabatic dispersion function is noted. Beyond $\gamma=1.16$, the extent of the dispersive zone starts shrinking as shown in Fig.~\ref{fig6}(d). This situation is similar to the case when only $\kappa/\mu$ was varied alone (shown in Fig.~\ref{fig2}(b)). For the case with $\kappa/\mu=0.25$, the range of $\eta$ for which the dispersion function may be approximated by that obtained from LEE remains almost invariant with change in the value of $\gamma$. For higher values of $\kappa/\mu$, the trend remains similar to as noted for $\kappa/\mu=0.25$, with the value of $\gamma$ corresponding to $\left(\eta_b\right)_{max}$ increasing. This feature would be illustrated later in Fig.~\ref{fig8}.

In Fig.~\ref{fig7}, we show the variation of the relative diffusion coefficient with depressed wavenumber $\eta$ when $\gamma$ is varied. When $\gamma=1.0$, the relative diffusion coefficient $\nu_3/\nu_1$ for $\kappa/\mu=0$ and $0.25$ is identical and invariant of $\eta$ as also predicted by Eq.~(\ref{eq26a}). Similarly, $\nu_{(4,5)}/\nu_1$ is constant in the dispersive zone before the corresponding bifurcation point $\eta_b$. The value of this constant relative diffusion coefficient is given by the corresponding value for $\alpha_*$ for $\kappa/\mu=0$ and $0.25$, respectively. So, it follows the trend $T_0$ as described in Sec.~\ref{effdiffbulk}. As $\gamma$ is increased the variation of relative diffusion coefficients show different trends for $\kappa/\mu=0$ and $0.25$ as shown in Fig.~\ref{fig7}(b,c). For the first case trend $T_1$ is followed which is shown in Fig.~\ref{fig7}(b) for $\kappa/\mu=0$ and $\gamma>1$. Here, $\nu_3/\nu_1$ monotonically increase with $\eta$ while being more than $1$ and $\nu_{(4,5)}/\nu_1$ monotonically decreases for $\eta<\eta_b$. For $\kappa/\mu=0.25$ and $\gamma>1$, trend $T_2$ is observed as shown in Fig.~\ref{fig7}(c). Here, $\nu_3/\nu_1$ monotonically decreases while that for the acoustic modes increase for $\eta<\eta_b$. For both the cases, the level of relative diffusion coefficient increases as $\gamma$ is increased. For even higher values of $\kappa/\mu$, the trend for the relative diffusion coefficients is similar to $T_2$.


Another factor appearing in the dispersion relationship is the Prandtl number $Pr=c_p\mu/\hat{k}$ which is a ratio of momentum and thermal diffusivity \cite{white1966,tritton2012}. Typical values of $Pr$ for common gases are tabulated in Table~\ref{tab1}. For common gases, $Pr$ is slightly less than $1$, while for most liquids it is generally much more than $1$ \cite{tritton2012}. A high Prandtl number indicates almost adiabatic fluid. For $Pr\to \infty$, Eq.~(\ref{eq12ab}) shows that $\lambda_3=0$ while $\lambda_{4,5}=-\alpha_*\pm \sqrt{\alpha_*^2-{1}/{\eta^2}}$. Therefore, in this limit the entropic mode simply convects without diffusion while the diffusion of the acoustic modes for $\eta<\eta_b$ are independent of $\eta$ and only depends on $\kappa/\mu$ as also noted through Eq.~(\ref{eq26a}). 


The variation of the dispersion function $\widehat{\mathcal{M}}_5$ with depressed wavenumber $\eta$ is for various values of $Pr$ ranging from $0.5$ to $1$ is similar to what is shown in Fig.~\ref{fig6}(c). For a given $\kappa/\mu$, there is a definite value of $Pr$ for which $\left(\eta_b\right)_{max}$ was noted. For a fixed $\kappa/\mu$ and $\gamma$, as $Pr$ is varied, the variation of the relative diffusion characteristic $\nu_j/\nu_1$ either follows $T_1$ or $T_2$ as shown in Fig.~\ref{fig7}(b) and \ref{fig7}(c), respectively. At lower $\kappa/\mu$, trend $T_1$ is noted while for higher values of $\kappa/\mu$ trend $T_2$ is noted. Intermediate between the trends $T_1$ and $T_2$, $T_0$ is noted at a particular value of $Pr=Pr_*$. The value of $Pr_*$ is noted to decrease with $\kappa/\mu$ for a fixed $\gamma$. We have noted for $\gamma=1.4$, $Pr_*=0.75$ and $0.6316$ for $\kappa/\mu=0$ and $0.25$, respectively. These results are note presented here for brevity.

\subsection{Estimation of $\kappa/\mu$ for significant disturbance evolution}

We have noted in Figs.~\ref{fig4} and \ref{fig5} that, relatively significant disturbance evolution is noted for the case which are close to the parameters for which $\eta_b$ is maximum. Following this principle, we next try to track the conditions for which $\eta_b$ is maximum for variations in $\gamma$ and $Pr$. This would give one idea regarding the range of parameters for which relatively significant disturbance (mostly acoustic) evolution may be noted in the physical plane. In Fig.~\ref{fig8}(a,b), we plot $\eta_b$ as a function of $\gamma$ and $Pr$, respectively for indicated values of bulk viscosity ratio $\kappa/\mu$. For Fig.~\ref{fig8}(a), $Pr=0.72$ while for Fig.~\ref{fig8}(b), $\gamma=1.4$ are used. One notes from Fig.~\ref{fig8}(a) that except for the case $\kappa/\mu=0$ where $\eta_b$ monotonically decreases with $\gamma$, there exists a particular value of $\gamma$ for which $\eta_b$ is maximum. This value of $\gamma$ is noted to increase with $\kappa/\mu$. Similarly, for fixed $\gamma$ and $\kappa/\mu$, there is a definite value of $Pr$ for which $\eta_b$ is maximum. We note that this particular value of $Pr$ for which $\eta_b$ is maximum, decreases with increase in $\kappa/\mu$.

It is to be noted that the discriminant for the cubic equation for $\eta_b^2$ \textit{i.e.}, Eq.~(\ref{eq15}) is negative. As discussed in Appendix-I, therefore, all the three roots are real (see Eq.~(\ref{Aeq5})), while one of these are the real physical one. In fact, these three real roots are the $x$-coordinates of three points which lie on a circle in the complex plane with radius $\hat{r}$ and center at $b/3$ as given by Eq.~(\ref{Aeq5}) in Appendix-I. In Fig.~\ref{fig8}(c), we plot $\log(\hat{r})$ for the cubic Eq.~(\ref{eq15}) as a function of $\kappa/\mu$ and $\gamma$ for $Pr=0.72$. Here, $\gamma$ is varied from $1$ to $2$, while $\kappa/\mu$ varies from $0$ to $1$. We note that
$\log(\hat{r})$ is maximum along the line denoted by $AB$ in Fig.~\ref{fig8}(c). The empirical equation of this line may be given as
$\kappa/\mu=1.3705\gamma - 1.302$. This line denotes the value of $\kappa/\mu$ for which $\eta_b$ is maximum for a given value of $\gamma$ for $Pr=0.72$. At $Pr=0.72$ and $\gamma=1.4$, $\kappa/\mu = 0.616$ for maximum $\eta_b$, which is almost same as that predicted by Fig.~\ref{fig2}(c). Similarly, when we plot $\log(\hat{r})$ in the $(\kappa/\mu,\;1/Pr)$-plane for $\gamma=1.4$ in Fig.~\ref{fig8}(d), we see $\log(\hat{r})$ is maximum along the line $CD$ whose equation is obtained as $\kappa/\mu=1.382/Pr-1.308$. In plotting Fig.~\ref{fig8}(d), we choose $1/Pr$ as the ordinate as we note that the Prandtl number appears in the dispersion relationship as $1/Pr$ (see Sec.~\ref{App_C} in Appendix-I). From the above equation one finds that at $Pr=0.72$ and $\gamma=1.4$, $\kappa/\mu=0.611$ for maximum $\eta_b$ which is also close to the value predicted in Fig.~\ref{fig2}(c). Combining these two equation, we suggest that for a given $\gamma$ and $Pr$, the value of $\kappa/\mu$ for maximum $\eta_b$ may be obtained empirically from

\begin{eqnarray}
  \frac{\kappa}{\mu}=2.657 \left(1.3705\gamma - 1.302\right) \left(\frac{1.382}{Pr}-1.308 \right)
\end{eqnarray}   

\noindent The implication of the above equation is that for a given $\gamma$ and $Pr$, one can predict the value of bulk viscosity ratio
$\kappa/\mu$ close to which relatively significant acoustic disturbance evolution may be noted as also shown in Sec.~\ref{linearized}.
This of course needs to be verified from detailed investigations using direct numerical simulations.

\section{Summary and conclusion} \label{sec_iii}

Here, the characteristics of the dispersion relation for 3D linearized compressible NSE is presented. The 3D compressible NSE supports five type of waves, two vortical, one entropic and two acoustic modes. The vortical modes are non-dispersive but diffusive in nature. The diffusion coefficient of the vortical mode is $\nu_{1,2}=-K^2/Re$ and therefore is independent of the Mach number. The entropic mode is also non-dispersive but diffusive in nature. In contrast, the acoustic modes are dispersive only up to a certain bifurcation wavenumber $K_b$. Beyond $K_b$ relative diffusion of one acoustic mode increases while that for the other decreases with wavenumber. If the non-dimensional mean flow is denoted as ${\bf c}=(c_x,c_y,c_z)$, then the dispersion relation is given as $\omega_j=\beta-i\Lambda_j$. Here, we denote the vortical modes by $j=1,2$, the entropic mode by $j=3$ and the two acoustic modes by $j=4,5$. The dispersion relation for the entropic and the acoustic modes are given by the cubic Eq.~(\ref{eq12a}). From the analysis presented here, we conclude the following points as enumerated below.

\begin{enumerate}
\item The relative diffusion coefficient for entropic and acoustic modes $\nu_{3,4,5}/\nu_1$ and dispersion function for acoustic modes $\widehat{\mathcal{M}}_{4,5}=\Lambda_{jI}M^2/Re$ only depends upon the depressed wavenumber $\eta=KM/Re$, the bulk viscosity ratio $\kappa/\mu$, specific heat ratio $\gamma$ and Prandtl number $Pr$ of the flow.
  
\item At lower wavenumber components, the deviation of the dispersion function from the inviscid and adiabatic case is proportional to $\eta^2$ at the leading order. The asymptotic expansion of the dispersion function and the relative diffusion coefficients of the entropic and acoustic modes for lower wavenumber components ($\eta<1$) is given in Sec.~\ref{ResD_B}. For long wavelength disturbances, the relative diffusion coefficients of the entropic and acoustic modes increase linearly with $\kappa/\mu$ and $\gamma$ while varying inversely with $Pr$.

\item When the bulk viscosity ratio is increased, the shape and extent of the dispersion function is altered significantly as described in Sec.~\ref{blkdisp}. The change is more significant for higher wavenumber components. We note that the depressed bifurcation wavenumber $\eta_b=K_bM/Re$ is maximum when $\kappa/\mu\simeq 0.6$ for $\gamma=1.4$ and $Pr=0.72$. Similar critical values of $\gamma$ and $Pr$ are also noted to exist for which $\eta_b$ is maximum. 

\item The relative diffusion coefficient for entropic and acoustic modes displays three types of trends depending upon $\kappa/\mu$, $\gamma$ and $Pr$, denoted here as $T_0$, $T_1$ and $T_2$. Following the trend $T_0$, the relative diffusion coefficients are independent of wavenumber. The trend $T_1$ corresponds to the case when $\nu_3/\nu_1$ monotonically increases while $\nu_{4,5}/\nu_1$ for $\eta<\eta_b$ monotonically decreases with $\eta$. The reverse scenario is noted for the trend $T_2$.
  
\item The smallest wavelength of the acoustic dispersive disturbances are noted to be of the similar order or one order higher than the mean-free path for most common dilute gases.

\item Lastly, we have empirically obtained a criterion on the bulk viscosity ratio $\kappa/\mu$ depending upon $\gamma$ and $Pr$ for which $\eta_b$ would be maximum. The significance of this criterion is that relatively significant evolution of acoustic and/or entropic disturbances are noted when the $\kappa/\mu$ is close to this critical value. This is established here considering linearized disturbance evolution for a given initial condition composed of entropic and acoustic disturbances as illustrated in Sec.~\ref{linearized}. The last aspect needs to be further verified from detailed investigations using direct numerical simulations.   

\end{enumerate}



\providecommand{\noopsort}[1]{}\providecommand{\singleletter}[1]{#1}%

\section{Appendix-I} \label{appendix} 

Here, the general solution solution of a cubic polynomial equation with real coefficients as given below is considered
for the ease of the reader. Such equations can be symbolically written as 
\begin{eqnarray}
  x^3+bx^2+cx+d=0 
  \label{Aeq1}
\end{eqnarray}

\noindent Following the substitution $y=\left(x-b/3\right)$ in the above equation, one gets the depressed cubic equation
\begin{eqnarray}
  y^3+py+q=0 
  \label{Aeq2}
\end{eqnarray}

\noindent where,
\begin{eqnarray}
  p=\left(c-\frac{b^2}{3}\right) \;\;\;\; \text{and} \;\;\;\; q=\left( \frac{2}{27}b^3-\frac{1}{3}bc+d\right) \nonumber
\end{eqnarray}

\noindent Following Vieta's substitution \cite{cubic} $y=\left(w-\frac{p}{3w}\right)$ to Eq.~(\ref{Aeq2}), one obtains

\begin{eqnarray}
  w^3-\frac{p^3}{27w^3}+q=0 
  \label{Aeq3}
\end{eqnarray}

\noindent Equation~(\ref{Aeq3}) is in essence a quadratic equation, which can be solved to obtain the solution for $w^3$ as
\begin{eqnarray}
  w^3=\left[-\frac{q}{2}\pm \sqrt{\Delta}\right] \nonumber
\end{eqnarray}

\noindent where, the discriminant $\Delta$ is given as
\begin{eqnarray}
  \Delta=\left(\frac{p^3}{27}+\frac{q^2}{4}\right) 
  \label{Aeq4}
  \end{eqnarray} 
\noindent Depending upon whether $\Delta$ is positive or negative or equal to zero, one obtains three different classes of the solution.

\subsection{Solution for $\Delta<0$}

As $\Delta<0$, one notes that $\Delta=-a^2$ where $a$ is a real positive number. Moreover, from Eq.~(\ref{Aeq4}) for this
case one also notes that $p<-s$, where $s=\sqrt[3]{\left({27q^2}/{4}\right)}>0$. Let, $R_1$ and $R_2$ represent the two solutions of the
Eq.~(\ref{Aeq4}) such that $R_1=\left((-q/2)+\sqrt{\Delta}\right)=Re^{i\psi}$ and $R_2=\left((-q/2)-\sqrt{\Delta}\right)=Re^{-i\psi}$ where,
$R=\sqrt{(-q/2)+a^2}$ and $\psi=\tan^{-1}\left(-2a/q\right)$. The general solution of the cubic equation Eq.~(\ref{Aeq1}) for
this case is given as
\begin{eqnarray}
  x_{1,2,3}=\frac{b}{3}+\left\{\begin{matrix}
2\hat{r} \cos\left(\frac{\psi}{3}\right) \\ 
2\hat{r} \cos\left(\frac{\psi}{3}+\frac{2\pi}{3}\right)\\ 
2\hat{r} \cos\left(\frac{\psi}{3}-\frac{2\pi}{3}\right)
  \end{matrix}\right.
  \label{Aeq5}
\end{eqnarray}
\noindent where $\hat{r}=\sqrt{-p/3}$. Here, all the three solutions are real. 

\subsection{Solution for $\Delta>0$} \label{App_B}

As $\Delta>0$, both $R_1$ and $R_2$ are real numbers. The general solution of the cubic equation Eq.~(\ref{Aeq1}) for
this case is given as
\begin{eqnarray}
  x_{1,2,3}=\frac{b}{3}+\left\{\begin{matrix}
u_1  \\ 
-\frac{1}{2} u_1 + i\frac{\sqrt{3}}{2} v_1 \\ 
-\frac{1}{2} u_1 - i\frac{\sqrt{3}}{2} v_1
  \end{matrix}\right.
  \label{Aeq6}
\end{eqnarray}
\noindent where $u_1=\left(\sqrt[3]{R_1}+\sqrt[3]{R_2}\right)$ and $v_1=\left(\sqrt[3]{R_1}-\sqrt[3]{R_2}\right)$. For this case, $x_1$ is real while $x_2$ and $x_3$ are complex conjugates.

\subsection{Solution for $\Delta=0$}

As $\Delta>0$, here $R_1=R_2=\left(-q/2\right)$ and hence, the general solution of the cubic equation Eq.~(\ref{Aeq1}) for
this case is given as

\begin{eqnarray}
  x_{1,2,3}=\frac{b}{3}+\left\{\begin{matrix}
2\sqrt[3]{\left(-q/2\right)}  \\ 
-\sqrt[3]{\left(-q/2\right)}  \\ 
-\sqrt[3]{\left(-q/2\right)} 
  \end{matrix}\right.
  \label{Aeq7}
\end{eqnarray}
\noindent
\noindent Here, all the roots are real with two identical roots corresponding to $x_2$ and $x_3$. 

\subsection{Connection with the cubic dispersion equation}
\label{App_C}

The cubic polynomial dispersion equation for the entropic and the two acoustic modes are given by Eq.~(\ref{eq12ab}). Comparing Eq.~(\ref{eq12ab}) with Eq.~(\ref{Aeq1}), one gets $b=C_8$, $c=\left(C_5 +{1}/{\eta^2} \right)$ and $d={C_7}/{\eta^2}$. From the general solution of the cubic polynomial equation we note the existence of three important parameters, \textit{viz.} $\left(-p/3\right)$, $\left(-q/2\right)$ and the discriminant $\Delta=\left(p^3/27+q^2/4\right)$, which in terms of $\eta$ are given as 
\begin{eqnarray}
  \left(-p/3\right)& = & a_0-\frac{1}{3}\left(\frac{1}{\eta^2}\right) \label{Aeq9} \\
  \left(-q/2\right) & = & b_0+b_1\left(\frac{1}{\eta^2}\right) \label{Aeq10} \\
  \Delta & = & d_0+d_1\left(\frac{1}{\eta^2}\right)+d_2\left(\frac{1}{\eta^4}\right)+\frac{1}{27}\left(\frac{1}{\eta^6}\right) \label{Aeq11}
\end{eqnarray}
\noindent where the above coefficients as functions of $\kappa/\mu$, $\gamma$ and $Pr$ are given as

$ \begin{aligned} 
&  a_0 & = & \left(\frac{1}{9}\right)\left(\frac{\kappa}{\mu}\right)^2 + \left(\frac{8}{27}\right)\left(\frac{\kappa}{\mu}\right)
    - \left(\frac{1}{9}\right)\left(\frac{\gamma}{Pr}\right)\left(\frac{\kappa}{\mu}\right)
    - \left(\frac{4}{27}\right)\left(\frac{\gamma}{Pr}\right)
    + \left(\frac{1}{9}\right)\left(\frac{\gamma}{Pr}\right)^2
    + \left(\frac{16}{81}\right) \nonumber 
\end{aligned} $

$ \begin{aligned} 
&  b_0 & = & - \left(\frac{16}{81}\right)\left(\frac{\kappa}{\mu}\right) - \left(\frac{4}{27}\right)\left(\frac{\kappa}{\mu}\right)^2
    - \left(\frac{1}{27}\right)\left(\frac{\kappa}{\mu}\right)^3 + \left(\frac{8}{81}\right)\left(\frac{\gamma}{Pr}\right)
    + \left(\frac{2}{27}\right)\left(\frac{\gamma}{Pr}\right)^2 \nonumber \\
&    & &
    - \left(\frac{1}{27}\right)\left(\frac{\gamma}{Pr}\right)^3
    + \left(\frac{1}{18}\right)\left(\frac{\gamma}{Pr}\right)\left(\frac{\kappa}{\mu}\right)^2
    + \left(\frac{1}{18}\right)\left(\frac{\gamma}{Pr}\right)^2\left(\frac{\kappa}{\mu}\right)
    + \left(\frac{4}{27}\right)\left(\frac{\gamma}{Pr}\right)\left(\frac{\kappa}{\mu}\right) - \left(\frac{64}{729}\right) \nonumber 
\end{aligned} $

$ \begin{aligned} 
&  b_1 & = & \left(\frac{2}{9}\right) + \left(\frac{1}{6}\right)\left(\frac{\kappa}{\mu}\right)
  - \left(\frac{1}{2}\right)\left(\frac{1}{Pr}\right) + \left(\frac{1}{6}\right)\left(\frac{\gamma}{Pr}\right) \nonumber 
\end{aligned} $

$ \begin{aligned} 
&  d_0 & = & - \left(\frac{64}{2187}\right)\left(\frac{\gamma}{Pr}\right)^2
    + \left(\frac{32}{729}\right)\left(\frac{\gamma}{Pr}\right)^3
    - \left(\frac{4}{243}\right)\left(\frac{\gamma}{Pr}\right)^4
    - \left(\frac{64}{729}\right)\left(\frac{\gamma}{Pr}\right)^2\left(\frac{\kappa}{\mu}\right) \nonumber \\
&    & & 
    + \left(\frac{8}{81}\right)\left(\frac{\gamma}{Pr}\right)^3\left(\frac{\kappa}{\mu}\right)
    - \left(\frac{2}{81}\right)\left(\frac{\gamma}{Pr}\right)^4\left(\frac{\kappa}{\mu}\right)
    - \left(\frac{8}{81}\right)\left(\frac{\gamma}{Pr}\right)^2\left(\frac{\kappa}{\mu}\right)^2
    - \left(\frac{4}{81}\right)\left(\frac{\gamma}{Pr}\right)^2\left(\frac{\kappa}{\mu}\right)^3 \nonumber \\
&    & & 
    - \left(\frac{1}{108}\right)\left(\frac{\gamma}{Pr}\right)^2\left(\frac{\kappa}{\mu}\right)^4
    + \left(\frac{2}{27}\right)\left(\frac{\gamma}{Pr}\right)^3\left(\frac{\kappa}{\mu}\right)^2
    + \left(\frac{1}{54}\right)\left(\frac{\gamma}{Pr}\right)^3\left(\frac{\kappa}{\mu}\right)^3
    - \left(\frac{1}{108}\right)\left(\frac{\gamma}{Pr}\right)^4\left(\frac{\kappa}{\mu}\right)^2 \nonumber 
\end{aligned} $

$ \begin{aligned} 
&  d_1 & = & + \left(\frac{1}{27}\right)\left(\frac{1}{Pr}\right)\left(\frac{\kappa}{\mu}\right)^3
    - \left(\frac{1}{54}\right)\left(\frac{\gamma}{Pr}\right)\left(\frac{\kappa}{\mu}\right)^3 
    + \left(\frac{2}{27}\right)\left(\frac{\gamma}{Pr}\right)^2\left(\frac{\kappa}{\mu}\right)^2 \nonumber \\
&    & & 
    - \left(\frac{1}{18}\right)\left(\frac{\gamma}{Pr}\right)\left(\frac{1}{Pr}\right)\left(\frac{\kappa}{\mu}\right)^2 
    - \left(\frac{2}{27}\right)\left(\frac{\gamma}{Pr}\right)\left(\frac{\kappa}{\mu}\right)^2
    + \left(\frac{4}{27}\right)\left(\frac{1}{Pr}\right)\left(\frac{\kappa}{\mu}\right)^2  \nonumber \\
&    & & 
    + \left(\frac{16}{81}\right)\left(\frac{1}{Pr}\right)\left(\frac{\kappa}{\mu}\right)
    - \left(\frac{8}{81}\right)\left(\frac{\gamma}{Pr}\right)\left(\frac{\kappa}{\mu}\right) 
    - \left(\frac{4}{27}\right)\left(\frac{\gamma}{Pr}\right)\left(\frac{1}{Pr}\right)\left(\frac{\kappa}{\mu}\right) \nonumber \\
&    & & 
    + \left(\frac{16}{81}\right)\left(\frac{\gamma}{Pr}\right)^2\left(\frac{\kappa}{\mu}\right)
    - \left(\frac{1}{18}\right)\left(\frac{\gamma}{Pr}\right)^2\left(\frac{1}{Pr}\right)\left(\frac{\kappa}{\mu}\right)
    - \left(\frac{1}{54}\right)\left(\frac{\gamma}{Pr}\right)^3\left(\frac{\kappa}{\mu}\right) \nonumber \\
&    & & 
    + \left(\frac{64}{729}\right)\left(\frac{1}{Pr}\right)
    - \left(\frac{32}{729}\right)\left(\frac{\gamma}{Pr}\right)
    + \left(\frac{32}{243}\right)\left(\frac{\gamma}{Pr}\right)^2
    - \left(\frac{8}{81}\right)\left(\frac{\gamma}{Pr}\right)\left(\frac{1}{Pr}\right) \nonumber \\
&    & & 
    - \left(\frac{2}{27}\right)\left(\frac{\gamma}{Pr}\right)^2\left(\frac{1}{Pr}\right)
    - \left(\frac{2}{81}\right)\left(\frac{\gamma}{Pr}\right)^3
    + \left(\frac{1}{27}\right)\left(\frac{\gamma}{Pr}\right)^3\left(\frac{1}{Pr}\right) \nonumber 
\end{aligned} $

$ \begin{aligned} 
&  d_2 & = & - \left(\frac{1}{108}\right)\left(\frac{\kappa}{\mu}\right)^2
    - \left(\frac{2}{81}\right)\left(\frac{\kappa}{\mu}\right)
    - \left(\frac{1}{6}\right)\left(\frac{1}{Pr}\right)\left(\frac{\kappa}{\mu}\right)
    + \left(\frac{5}{54}\right)\left(\frac{\gamma}{Pr}\right)\left(\frac{\kappa}{\mu}\right) \nonumber \\
&    & & 
    + \left(\frac{1}{4}\right)\left(\frac{1}{Pr}\right)^2
    - \left(\frac{1}{6}\right)\left(\frac{\gamma}{Pr}\right)\left(\frac{1}{Pr}\right)
    - \left(\frac{1}{108}\right)\left(\frac{\gamma}{Pr}\right)^2
    - \left(\frac{2}{9}\right)\left(\frac{1}{Pr}\right) \nonumber \\
&    & & 
    + \left(\frac{10}{81}\right)\left(\frac{\gamma}{Pr}\right)
    - \left(\frac{4}{243}\right) \nonumber 
\end{aligned} $

\end{document}